\documentclass[aps,unsortedaddress,groupedaddress,nofootinbib]{revtex4}

\usepackage{graphics,graphicx,epsfig}
\usepackage{epstopdf}
\usepackage{times}
\usepackage{natbib}
\usepackage{amsmath}
\usepackage{amssymb}
\usepackage{hyperref}
\usepackage{bm}
\usepackage{color}
\usepackage[normalem]{ulem}

\def\Ni{$^{56}\textup{Ni}$ }

\begin{document}

\title{Type Ia supernovae, standardisable candles, and gravity}

\author{Bill S. Wright$^{1,2}$}
\email{bill.wright@port.ac.uk}
\author{Baojiu Li$^1$}
\email{baojiu.li@durham.ac.uk}
\affiliation{$^1$Institute for Computational Cosmology, Department of Physics, Durham University, Durham DH1 3LE, UK\\
$^2$Institute of Cosmology \& Gravitation, University of Portsmouth, Portsmouth, Hampshire, PO1 3FX, UK}

\begin{abstract}
Type Ia supernovae (SNe Ia) are generally accepted to act as standardisable candles, and their use in cosmology led to the first confirmation of the as yet unexplained accelerated cosmic expansion. Many of the theoretical models to explain the cosmic acceleration assume modifications to Einsteinian General Relativity which accelerate the expansion, but the question of whether such modifications also affect the ability of SNe Ia to be standardisable candles has rarely been addressed. This paper is an attempt to answer this question. For this we adopt a semi-analytical model to calculate SNe Ia light curves in non-standard gravity. We use this model to show that the average rescaled intrinsic peak luminosity -- a quantity that is assumed to be constant with redshift in standard analyses of Type Ia supernova (SN Ia) cosmology data -- depends on the strength of gravity in the supernova's local environment because the latter determines the Chandrasekhar mass -- the mass of the SN Ia's white dwarf progenitor right before the explosion. This means that SNe Ia are no longer standardisable candles in scenarios where the strength of gravity evolves over time, and therefore the cosmology implied by the existing SN Ia data will be different when analysed in the context of such models. As an example, we show that the observational SN Ia cosmology data can be fitted with both a model where $(\Omega_{\rm M}, \Omega_{\Lambda})=(0.62, 0.38)$ and Newton's constant $G$ varies as $G(z)=G_0(1+z)^{-1/4}$ and the standard model where $(\Omega_{\rm M}, \Omega_{\Lambda})=(0.3, 0.7)$ and $G$ is constant, when the Universe is assumed to be flat.
\end{abstract}

\maketitle

\section{Introduction}\label{sec:intro}

Cosmology observations from Type Ia supernovae (SNe Ia) gave the first evidence for a late-time acceleration in the expansion of the Universe \citep{Perlmutter1999,Riess1998}. A Type Ia supernova (SN Ia) is the cataclysmic explosion of a white dwarf star that occurs when the white dwarf accretes enough mass from a binary partner for the material in its core to undergo runaway thermonuclear fusion. SNe Ia are thought to act as standardisable candles because of an observed relationship between a SN Ia's peak brightness and how rapidly this peak brightness is achieved and subsequently left behind \cite{Phillips1993} (the so-called width-luminosity relation, or WLR). After standardisation procedures, which are often at least partially based on the WLR, are applied, any remaining difference in the peak brightnesses of the two SNe Ia should be due to a difference in distance to the observer. Thus the relative distances between SNe Ia can be measured, and along with measurements of their redshifts can be used to infer the details of the expansion of the Universe through the construction of the distance-redshift relation.

After the late-time acceleration of the expansion of the Universe was discovered, Einstein's idea of a small, positive cosmological constant $\Lambda$ was revived and established as the leading candidate for the acceleration's origin. However, the idea of $\Lambda$ as the cause of cosmic acceleration is not without theoretical difficulties, such as the cosmological constant fine-tuning and coincidence problems \cite{cosmoconstprob, coincidenceprob}. These problems have motivated a wide array of theories of dynamical dark energy \cite{Copelandetal2006} or modified gravity \cite{Joyceetal2015, Koyama2016} which aim to explain the smallness of $\Lambda$ (or its substitute) using dynamical or more natural mechanisms. The latter class of models has received growing interest in recent years, partly because current and next-generation cosmological surveys (such as e{\sc boss} \cite{eBOSS}, {\sc des} \cite{DES}, {\sc hsc} \cite{HSC}, {\sc desi} \cite{DESI}, {\sc lsst} \cite{LSST}, {\sc Euclid} \cite{Euclid}, {\sc 4most} \cite{4MOST}, {\sc wfirst} \cite{WFIRST} and {\sc ska} \cite{SKA}) will allow their theoretical predictions to be confronted with precision data. The use of Einstein's General Relativity (GR) as the foundation of modern cosmology is a vast extrapolation of its validity beyond the length and energy scales at which it has been rigorously tested \cite{Will2014}, and thus testing the validity of GR with unprecedented precision in this new regime of cosmological scales is a vital task.

An unintended and less well recognised consequence of introducing theories of modified gravity is their potential impact on the astrophysics of SNe Ia themselves -- in particular, their ability to act as standardisable candles could be affected. A fairly common feature of these theories is that the strength of gravity varies over cosmic time and/or across space, as a result of which some key properties of the white dwarf progenitors of the SNe Ia, such as their mass, can become redshift dependent. The redshift dependence of these key properties may in turn affect the intrinsic peak luminosities of the SNe Ia. If this is the case, then the measurement of the acceleration such theories are introduced to explain might need to be reinterpreted as the very result of their introduction. In the extreme case, modified gravity theories may produce an acceleration that is no longer supported by the SN Ia data once the data is reinterpreted in the context of the new theory. Evidently, this is an interesting question that is not just important for consistency in the study and testing of modified gravity theories, but also relevant to the general cosmological community. It has long been recognised that any evolution of the SNe Ia intrinsic luminosity $L$ with redshift would affect the distances measured and therefore the values of the cosmological parameters deduced from SN Ia cosmology observations \cite{Drelletal2000}. There have also been some earlier efforts to relate this evolution of intrinsic luminosity to an evolution of the strength of gravity through the value of Newton's gravitational constant $G$ \cite{Amendolaetal1999,Garcia-Berroetal1999,RiazueloUzan2002}. However, these initial studies assumed a straightforward proportionality between $L$ and $G$, and either did not attempt to produce light curves \cite{Amendolaetal1999,Garcia-Berroetal1999}, or if they did produce light curves \cite{RiazueloUzan2002} did not attempt to reproduce the WLR or quantitatively verify that the standardisation procedures still work when $G \neq G_0$, where $G_0$ is the value of $G$ at the present day. This method has also been utilised to place constraints on the variation of $G$ using the observational dispersion in SNe Ia absolute magnitudes \cite{Gaztanagaetal2002,LorenAguilaretal2003,MouldUddin2014}.

In this paper we further develop these early works and introduce an alternative method to treat SNe Ia in modified gravity theories that allows us to produce SNe Ia light curves in order to verify whether the WLR is reproduced in non-standard gravity, and then use this information to tackle the issue of whether the ability of SNe Ia to act as standardisable candles is affected.

Modeling the impact of modified gravity on SN Ia astrophysics is a highly nontrivial task. The pre and post-collapse phases of the evolution of SNe Ia are both typical astrophysical laboratories where all four types of fundamental interactions play a role and the many physical processes going on are not yet fully understood. Even with the current best knowledge of these physical processes, the SN Ia evolution cannot be accurately followed without expensive hydrodynamical simulations. This is further complicated by the strong variations displayed in the physical properties of the SN Ia population, such as their burning conditions and the $^{56}$Ni mass produced in the thermonuclear reactions, the latter being a critical quantity determining the intrinsic SN Ia luminosity. If we add modifications to gravity on top of all these, the situation only becomes worse. Clearly, a simplified approach is needed for initial proof-of-concept studies before embarking on a full numerical investigation.

Our approach makes significant simplification of the study in three ways: firstly the use of a semi-analytical model for SNe Ia light curves that has been demonstrated to work quite well in explaining the observed behaviours of the light curves; secondly the treatment of modified gravity as a time variation of Newton's constant $G$, which affects SN Ia astrophysics mainly by modifying the mass of the white dwarf progenitor -- the Chandrasekhar mass; and thirdly the use of a simplified standardisation procedure (in comparison to those used for observational work \cite{MLCS,SALT2}) that involves rescaling light curves so that their shape around the peak matches that of a template. We will show that, using this simplified procedure, the semi-analytical light curve model can successfully reproduce the WLR and the standardisability of SNe Ia light curves in the case of a redshift-independent value for Newton's gravitational constant $G(z)=G_0$, which corresponds to the well-known result of the Chandrasekhar mass $M_{\rm Ch}=1.44M_\odot$. However, if $M_{\rm Ch}$ takes different values at different redshifts due to $G(z)\neq G_0$, we will show that although the WLR is reproduced for $G \neq G_0$, our procedure for the standardisation of SNe Ia light curves based on this WLR gives rise to different rescaled intrinsic peak luminosities than predicted by $G=G_0$, suggesting that SNe Ia are no longer conventional standardisable candles with the same rescaled intrinsic peak luminosities at all redshifts.

We will present a simple numerical example that utilises this result and demonstrates that the same SN Ia data can be fitted by two cosmological models: one with $(\Omega_{\rm M},\Omega_{\Lambda})=(0.3,0.7)$ and $G=G_0$, and the other with $(\Omega_{\rm M},\Omega_{\Lambda})=(0.62,0.38)$ and $G(z)=G_0(1+z)^{-1/4}$, where $z$ is the redshift, and $\Omega_{\rm M}$ and $\Omega_{\Lambda}$ are respectively the present-day density parameters for non-relativistic matter and the cosmological constant $\Lambda$.

This paper is organised as follows. We start in Section \ref{ssec:LCM} by describing a model capable of producing the light curves of SNe Ia for a given set of input parameters, show in Section \ref{ssec:LCparams} how varying each of these input parameters affects the light curve, discuss the relationship between two of these input parameters, nickel-56 mass $M_{\rm Ni}$ and ejecta opacity $\kappa$, that is essential to the rescaling of SNe Ia light curves in Section \ref{ssec:MNikappa}, and then in Section \ref{ssec:GdepLCM} deduce how modifications to gravity would affect the values of the input parameters and therefore the light curves. Once we have this gravity-dependent light curve model, we use it to investigate how the results of the light curve rescaling process change under modified gravity in Section \ref{ssec:GDepWLR}, before presenting the numerical example mentioned above in Section \ref{ssec:NumEx}. Finally, we conclude in Section \ref{sec:conc}.

\section{SN Ia Astrophysics}\label{sec:SN Iaastro}

To construct a model capable of capturing the effect of a time-dependent local strength of gravity on a SN Ia light curve, we first identify a model that can reproduce a light curve for a given set of input parameters in Section \ref{ssec:LCM}, investigate how the light curve depends on each parameter in Section \ref{ssec:LCparams}, discuss a key component of the model that allows SNe Ia light curves to be standardisable in Section \ref{ssec:MNikappa}, and then determine how the values of those parameters depend on the value of $G$ in Section \ref{ssec:GdepLCM}.

\subsection{Light curve model}
\label{ssec:LCM}

\subsubsection{Physics of the light curve} \label{sssec:equation}
A SN Ia is triggered when a carbon/oxygen white dwarf accretes enough mass from a binary partner to increase the temperature and density in its core past the level required to restart nuclear fusion. This initial fusion begins a runaway thermonuclear process known as carbon detonation that releases enough energy to destroy the white dwarf. During this process, large quantities of the radioactive isotope nickel-56 ($^{56}\textup{Ni}$) are produced. The $^{56}\textup{Ni}$ undergoes positron decay to cobalt-56 ($^{56}\textup{Co}$) which in turn decays via positron emission to the stable isotope iron-56 ($^{56}\textup{Fe}$) \citep{Colgate1969}. The decay chain is simplified as follows:
\begin{equation}
{}^{56}_{28}\textup{Ni} \to {}^{56}_{27}\textup{Co} + {}^{\ 0}_{+1}\mathrm{e}^+ + \gamma \to {}^{56}_{26}\textup{Fe} + 2\ {}^{\ 0}_{+1}\mathrm{e}^+ + \gamma.
\end{equation}
The radiation produced in these decays is in the form of short-wavelength gamma rays. Throughout this paper we will treat such gamma rays as unobservable and only consider the SN Ia's ultraviolet+optical+infrared (UVOIR) light curve. Therefore to contribute to the light curve, the radiation produced in these decays must first increase its wavelength through one of many possible interactions with the supernova ejecta material thrown out in the initial explosion of the white dwarf progenitor. This longer wavelength radiation can then diffuse through the supernova ejecta and be observed. Gamma rays that diffuse through the ejecta without interacting will not contribute to the UVOIR light curve -- this is known as gamma ray leakage.

\begin{figure*}
\begin{center}
\includegraphics[width=\textwidth]
{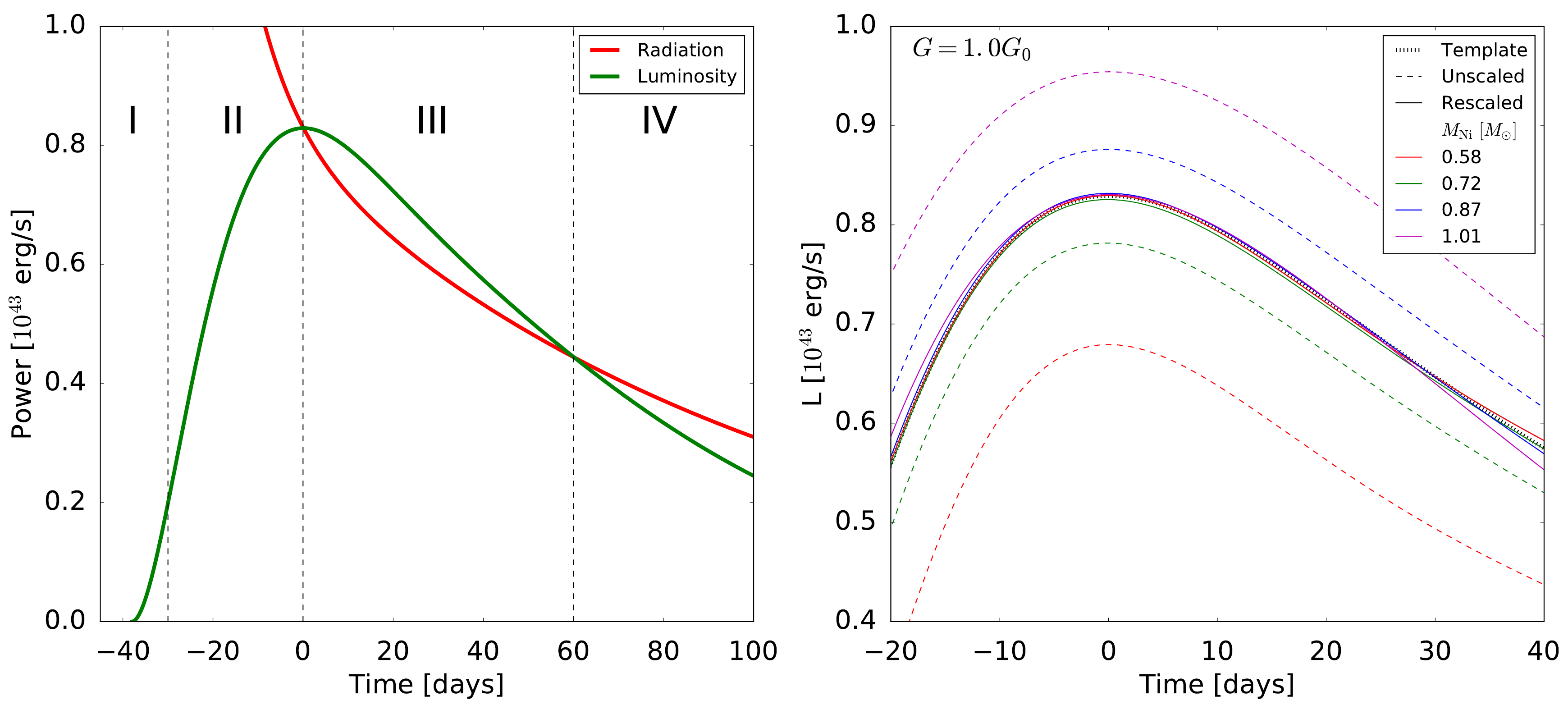}
\caption[]{{\it Left panel}: Sketch of a typical UVOIR light curve for a SN Ia, along with the power produced by the radioactive decay chain of $^{56}\rm Ni$. The Roman numerals correspond to the phases described in the text. {\it Right panel}: Rescaling of light curves from a SN Ia population with varying $M_{\mathrm{Ni}}$ in standard gravity ($G=G_0$). A template curve, whose shape around the peak the other curves have been rescaled to match with, is also shown.}
\label{fig:G0_Rescaling}
\vspace{-3ex}
\end{center}
\end{figure*}

A qualitative description of each phase of the diffusion process for the post-interaction, longer wavelength radiation is given below and is accompanied by a sketch of the corresponding UVOIR light curve in the left panel of Fig.~\ref{fig:G0_Rescaling}:

\begin{enumerate}
	\item[I:] At early times, the outer layers of the supernova ejecta are still hot and densely packed, with high opacity to radiation of all wavelengths. Thus at this stage the instantaneous luminosity observed from the supernova is only a small fraction of the instantaneous power from radioactive decay in the centre of the ejecta, and this can be seen in the small initial brightness of the supernova's light curve.
	\item[II:] Gradually, as the ejecta expands and disperses, its opacity to longer wavelength radiation falls and the amount of UVOIR radiation that can escape increases, until the ejecta becomes essentially translucent and UVOIR radiation can escape unimpeded. This can be seen as the light curve steadily rises to its peak after the initial explosion.
	\item[III:] Once the ejecta has become essentially fully translucent to UVOIR wavelengths, the trapped UVOIR radiation that had been produced at earlier times before the ejecta became translucent can escape. This results in the instantaneous observed luminosity temporarily rising above the instantaneous power from radioactive decay until the excess trapped UVOIR radiation energy has escaped. This can be seen in the light curve shortly after the time of peak brightness.
	\item[IV:] After this point, the observed UVOIR luminosity falls below the instantaneous power from radioactive decay. This is because the opacity of the ejecta at short wavelengths is now small enough that a significant fraction of the radiation produced leaks out as unobserved gamma rays without interacting to become longer wavelength UVOIR radiation, and so does not contribute to the UVOIR light curve.
\end{enumerate}

In order to calculate the light curve produced by the supernova in this complex situation, the radiative transport equations for the decay radiation propagating through the ejecta must be solved. This problem has been tackled by many groups using large numerical simulations, for example see Refs.~\cite{Blinnikov2006} and \cite{Kasen2006}. However, such calculations are computationally intensive and are beyond the scope of this research. Instead, a semi-analytical treatment of an approximated version of the full problem is used, a method which has shown success in reproducing the standard observed behaviour of SNe Ia and the results of more complex numerical simulations \citep{Jeffery1999, Pinto2000a}. The method that follows is based on the treatments in Refs.~\cite{Arnett1980}, \cite{Arnett1982}, and \cite{Chatzopoulos2012} (henceforth known as A80, A82, and C12 respectively). A condensed derivation of the equation for the light curve is presented here, but a fully detailed version can be seen in Appendix \ref{App:LCDerivation}. The following assumptions and approximations are made:

\begin{enumerate}
	\item Spherically symmetric system;
	\item Homologous expansion of ejecta -- see Eq.~(\ref{eq:homexp1});
	\item Ejecta gas that is dominated by radiation pressure;
	\item Diffusion approximation -- optically thick ejecta (optical depth $>$ 1);
	\item Constant effective opacity of ejecta;
	\item Radioactive decay as the only source of energy;
	\item Concentrated distribution of $^{56}\textup{Ni}$ in the centre of the system.
\end{enumerate}

We start by applying the first law of thermodynamics to the expanding supernova ejecta:
\begin{equation}
\dot{E} + P \dot{V} = -\frac{\partial L}{\partial m} + \epsilon~,
\label{eq:1stlaw1}
\end{equation}
where $E=\alpha T^4 V$ is the specific energy, $P=\alpha T^4/3$ is the pressure, $T$ is the temperature of the ejecta, $V=1/\rho$ is the specific volume, $\rho$ is the density of the ejecta, $\alpha=4\sigma /c$ is the radiation constant, $\sigma$ is the Stefan-Boltzmann constant, the $\dot{y}$ notation represents the partial derivative with respect to time $\partial y/\partial t$,  $L$ is the luminosity output of the system, $m$ is the mass, and $\epsilon$ is the rate of energy per unit mass added to the system. The first term, $\dot{E}$, is the rate of change in energy density, and $P\dot{V}$ represents the specific work involved in expanding the ejecta, so the equation shows that the sum of the rate of change in energy density and the specific work are equal to the sum of the energy per unit mass added to the system (positive) and the luminosity output of the system per unit mass (negative).

The source of energy in this system, $\epsilon$, is the radioactive decay of $^{56}\textup{Ni}$ to $^{56}\textup{Co}$ and the subsequent decay of $^{56}\textup{Co}$ to stable $^{56}\textup{\textup{Fe}}$, and is given by
\begin{equation}
\epsilon(r, t) = \xi(r) \left[ (\epsilon_{\textup{Ni}}-\epsilon_{\textup{Co}}) \mathrm{e}^{-t/\tau_{\textup{Ni}}} + \epsilon_{\textup{Co}} \mathrm{e}^{-t/\tau_{\textup{Co}}} \right]~,
\label{eq:eps41}
\end{equation}

\noindent
where $\xi(r)$ is the radial distribution of $^{56}\textup{Ni}$ in the ejecta, $\epsilon_{\textup{Ni}}\mathrm{=3.9\times 10^{10}}$ $\mathrm{erg/s/g}$ and $\epsilon_{\textup{Co}} \rm = 3.9 \times 10^{10} ~erg/s/g$ are the energy generation rates from $^{56}\textup{Ni}$ and $^{56}\textup{Co}$ decays respectively, and $\tau_{\textup{Ni}}\mathrm{=8.8~days}$ and $\tau_{\textup{Co}}\mathrm{=111.3~days}$ are the lifetimes of  $^{56}\textup{Ni}$ and $^{56}\textup{Co}$ respectively \citep{Nadyozhin1994}.

In the diffusion approximation, the luminosity of a shell of the ejecta at radius $r$ is related to the temperature of that shell by
\begin{equation}
L = -4\pi r^2 \frac{\Gamma c a}{3} \frac{\partial T^4}{\partial r}~,
\label{eq:luminosity1}
\end{equation}

\noindent
where $\Gamma=1/\rho \kappa$ is the mean free path in the ejecta, $\kappa$ is the effective opacity of the ejecta, and $c$ is the speed of light. The temperature can be expressed as a separation of variables:
\begin{equation}
T(r,t)^4=\psi(r) \phi(t) T^4_{00} {\left[\frac{R_0}{R(t)}\right]}^4~,
\label{eq:temp1}
\end{equation}

\noindent
where the temperature's radial dependence is contained in $\psi(r)$, and its time dependence in $\phi(t)$. $T_{00}$ is the initial temperature at zero radius. Shortly after the initial supernova explosion, the expansion of the ejecta should become homologous such that the radial extent of the surface of the ejecta at time $t$, $R(t)$, is  given by
\begin{equation}
R(t) = R_0 + v_{\mathrm{sc}}t~,
\label{eq:homexp1}
\end{equation}

\noindent
where $R(t)$ has advanced constantly at a scale velocity $v_{\mathrm{sc}}$ from its initial position at shock breakout $R_0$. For a SN Ia, this can be taken as the radius of the white dwarf progenitor \cite{Piroetal2009}.

Solving Eq.~(\ref{eq:1stlaw1}) using Eqs.~(\ref{eq:luminosity1})-(\ref{eq:homexp1}) (see Appendix \ref{App:LCDerivation} for details) gives the surface luminosity as a function of time, which is an equation for the light curve, as
\begin{widetext}
\begin{multline}
L_{\mathrm{surf}}(t)=\frac{2M_{\textup{Ni}}}{\tau_{\mathrm{m}}} \mathrm{e}^{ -\left(\frac{2 R_0 t}{v_{\mathrm{sc}} \tau^2_m} + \frac{t^2}{\tau^2_m}\right) }
\bigg[ (\epsilon_{\textup{Ni}}-\epsilon_{\textup{Co}}) \int^{t}_{0} \left(\frac{R_0}{v_{\mathrm{sc}}\tau_{\mathrm{m}}} + \frac{t^{\prime}}{\tau_{\mathrm{m}}}\right) \mathrm{e}^{\left( \frac{t^{\prime 2}}{\tau^2_m} + \frac{2R_0 t^{\prime}}{v_{\mathrm{sc}}\tau^2_m} \right)} \mathrm{e}^{-t^{\prime}/\tau_{\textup{Ni}}} {\rm d}t^{\prime} \\
 + \epsilon_{\textup{Co}} \int^{t}_{0} \left(\frac{R_0}{v_{\mathrm{sc}}\tau_{\mathrm{m}}} + \frac{t^{\prime}}{\tau_{\mathrm{m}}}\right) \mathrm{e}^{\left( \frac{t^{\prime 2}}{\tau^2_m} + \frac{2R_0 t^{\prime}}{v_{\mathrm{sc}}\tau^2_m} \right)} \mathrm{e}^{-t^{\prime}/\tau_{\textup{Co}}} {\rm d}t^{\prime} \bigg]~,
\label{eq:luminosity31}
\end{multline}
\end{widetext}

\noindent
where $M_{\textup{Ni}}$ is the initial mass of $^{56}\textup{Ni}$ in the ejecta, $\tau_{\mathrm{m}}=(2\kappa M_{\mathrm{ej}} /v_{\mathrm{sc}}\beta c)^{1/2}$ is the light curve timescale which determines how quickly the brightness rises to a peak and falls away again, $M_{\mathrm{ej}}$ is the total ejecta mass, and $\beta$ is a constant that depends on the ejecta's density profile. A80 calculates that a good approximation for a variety of density profiles is $\beta =13.8$, and this value is adopted here as well. Eq.~(\ref{eq:luminosity31}) will be completed by an overall time-dependent factor to account for gamma ray leakage -- see Appendix \ref{App:LCDerivation} for more details.

\subsubsection{Model parameters}\label{sssec:params}

It would appear the variables that need to be known in order to calculate the light curve using Eq.~(\ref{eq:luminosity31}) are the initial radius of shock breakout $R_0$, scale velocity $v_{\mathrm{sc}}$, initial $^{56}\textup{Ni}$ mass $M_{\textup{Ni}}$, effective opacity $\kappa$, and total ejecta mass $M_{\mathrm{ej}}$. However, $v_{\mathrm{sc}}$ can be calculated from the energetics of the supernova explosion. The kinetic energy $E_{\mathrm{K}}$ is calculated as the difference between the energy produced by nuclear fusion $E_{\mathrm{N}}$ and the gravitational binding energy $E_{\mathrm{G}}$ of the white dwarf progenitor \cite{Howell2006, Maeda2009}. The equation for $E_{\mathrm{N}}$ given in Ref.~\cite{Maeda2009} is
\begin{equation}\label{S12 4}
E_{\mathrm{N}} =[1.74f_{\textup{\textup{Fe}}}+1.56f_{\textup{Ni}}+1.24f_{\textup{Si}}] \Big( \frac{M_{\mathrm{ej}}}{M_{\odot}} \Big) \times 10^{51} \mathrm{~erg}~,
\end{equation}
where $f_{\textup{Fe}, \textup{Ni}, \textup{Si}, \textup{C/O}}$ are the initial fractions of the ejecta mass in the form of stable $^{56}\textup{Fe}$; radioactive $^{56}\textup{Ni}$; intermediate mass elements such as $\textup{Si}$, $\textup{Mg}$, and $\textup{S}$; and unburned carbon/oxygen respectively, and the fractions are governed by the relationship $f_{\textup{C/O}} = 1-f_{\textup{Fe}}-f_{\textup{Ni}}-f_{\textup{Si}}$. $f_{\textup{Ni}}$ can be rewritten in terms of the initial mass of $^{56}\textup{Ni}$ in the ejecta using $M_{\textup{Ni}}=f_{\textup{Ni}}M_{\mathrm{ej}}$. An empirical formula for $E_{\mathrm{G}}$ is prescribed in Ref.~\cite{Yoon2005}:
\begin{widetext}
\begin{align}\label{YL05 34}
E_{\mathrm{G}}(\rho_{\mathrm{c}}) = - \bigg[ 32.759747 + 6.7179802 \log_{10} \rho_{\mathrm{c}} - 0.28717609(\log_{10} \rho_{\mathrm{c}})^2 \bigg] \times 10^{50} \mathrm{~erg}~,
\end{align}
\end{widetext}
where $\rho_{\mathrm{c}}$ is the central density of the white dwarf progenitor. This allows the scale velocity $v_{\mathrm{sc}}$ to be calculated using: $E_{\mathrm{K}}=6M_{\mathrm{ej}}\ v_{\mathrm{sc}}^2=\left| E_{\mathrm{G}}-E_{\mathrm{N}} \right|$ \cite{Scalzo2012}. Thus the free parameter $v_{\mathrm{sc}}$ has been removed, but only at the expense of adding the mass fractions $f_{\textup{Fe}, \textup{Si}, \textup{C/O}}$ and the central density of the white dwarf progenitor $\rho_{\mathrm{c}}$ as free parameters instead.

At this stage the free parameters required are $R_0$, $M_{\mathrm{ej}}$, $\kappa$, $\rho_{\mathrm{c}}$, $M_{\textup{Ni}}$, and two out of three of $f_{\textup{Fe}, \textup{Si}, \textup{C/O}}$ since the fractions must sum to unity. However, this number can be reduced further by investigating the effects of varying the central density of the white dwarf progenitor on the ratio of yields of $^{56}\textup{Ni}$ and $^{56}\textup{Fe}$ produced in the initial supernova explosion, which leads to the following relationship \cite{Krueger2010}:
\begin{equation}\label{S12 6}
\frac{f_{\textup{Ni}}}{f_{\textup{Ni}} + f_{\textup{Fe}}} = 0.95 - 0.05 \times \frac{\rho_{\mathrm{c}}}{10^9 \mathrm{~g/cm^{3}}}~,
\end{equation}
such that $f_{\textup{Fe}}$ can be calculated provided $f_{\textup{Ni}}=M_{\rm Ni}/M_{\rm ej}$ and $\rho_{\mathrm{c}}$ are known, which removes $f_{\textup{Fe}}$ as a free parameter in our model. Thus only one of the three element fractions now need to be specified. There are strong limits on $f_{\textup{C/O}}$ in that it is well know that the amount of the unburned carbon and oxygen is very small, between $0-5\%$ of the total white dwarf mass \citep{Thomas2011}. For this reason, we choose $f_{\textup{C/O}}$ as the free parameter instead of $f_{\rm Si}$.

Finally, there is a relationship between the initial mass of \Ni in the ejecta $M_{\textup{Ni}}$ and the mean effective opacity of the ejecta $\kappa$ which allows us to eliminate $\kappa$ as an input parameter in our light curve model. Because knowledge of this relationship is vital to understanding the WLR observed in SNe Ia populations, the specifics of the $M_{\textup{Ni}}$-$\kappa$ relationship are discussed in greater detail in its own, dedicated section; Section \ref{ssec:MNikappa}.

Thus a semi-analytical model to calculate the light curve at all times can be created, based on Eq.~(\ref{eq:luminosity31}), as long as values for the following five parameters are specified: $M_{\mathrm{ej}}$, $M_{\textup{Ni}}$, $f_{\textup{C/O}}$, $\rho_{\mathrm{c}}$, and $R_0$. Although the total number of input parameters has not decreased from our initial set, this new set of input parameters can be estimated with much better justification, or previous work has established constraints on them as well.

\subsection{Dependence of model on parameters}
\label{ssec:LCparams}

Before moving on, we would like to test the model's dependency on each of the input parameters in order to gain an intuitive sense of how the model works, and to verify that the model behaves sensibly given our knowledge of the underlying supernova astrophysics.

\begin{figure*}
\includegraphics[width=\textwidth]{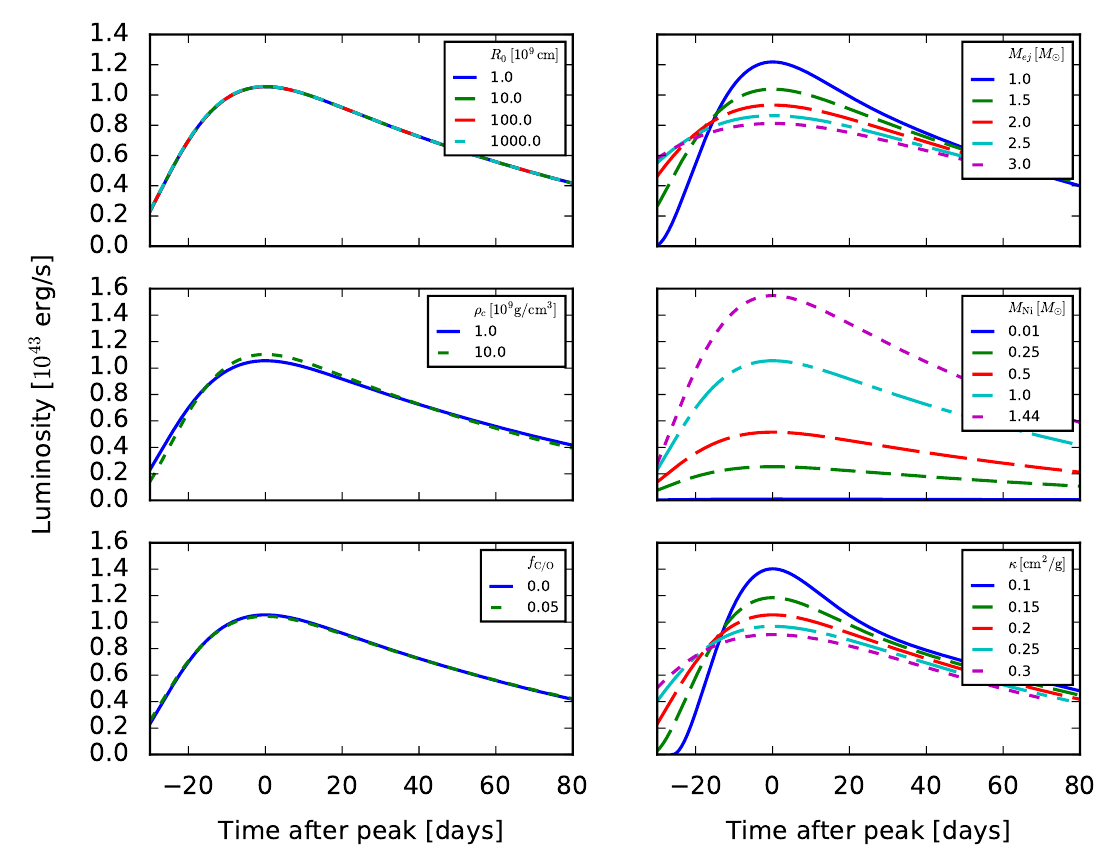}
\caption{Effect of varying each model parameter on the SN Ia light curve while the other parameters remain fixed. When not being varied, the parameters are fixed with values $R_0\mathrm{=10^{9}~cm}$, $M_{\mathrm{Ni}}\mathrm{=1.0}M_{\odot}$, $f_{\rm C/O}\mathrm{=0.0}$, $\rho_{\mathrm{c}}\mathrm{=1.0\times 10^9~g/cm^{3}}$, $M_{\mathrm{ej}}\mathrm{=1.44}M_{\odot}$, and $\kappa \mathrm{=0.2\ cm^2/g}$} \label{fig:param_dep}
\end{figure*}

In order to test the effect of model parameters on the light curve, each parameter was varied over an observationally motivated (or allowed) range in turn, while the other parameters were held constant. For now we assume $M_{\textup{Ni}}$ and $\kappa$ are independent input parameters, although we will show in Section \ref{ssec:MNikappa} that there is a relationship between the two. Further, we neglect the minor effect of varying white dwarf mass on the binding energy, though this will be included later. The results are shown in Fig.~\ref{fig:param_dep}.

Given that the value of the radius of initial shock breakout can be taken to be the radius of the white dwarf progenitor \cite{Piroetal2009}, the values of $R_0$ would be expected to be in the range $R_0\mathrm{=10^{7}-10^{9}~cm}$. In the upper left panel of Fig.~\ref{fig:param_dep} we can see that varying $R_0$ within this range yields essentially identical light curves. This corresponds to the limit $R_0 \rightarrow 0$ in Eq.~(\ref{eq:luminosity31}) which, as discussed in Ref.~\cite{Arnett1982}, means that the light curve model essentially no longer depends on $R_0$.

\begin{figure*}
\begin{center}
\includegraphics[width=\textwidth]{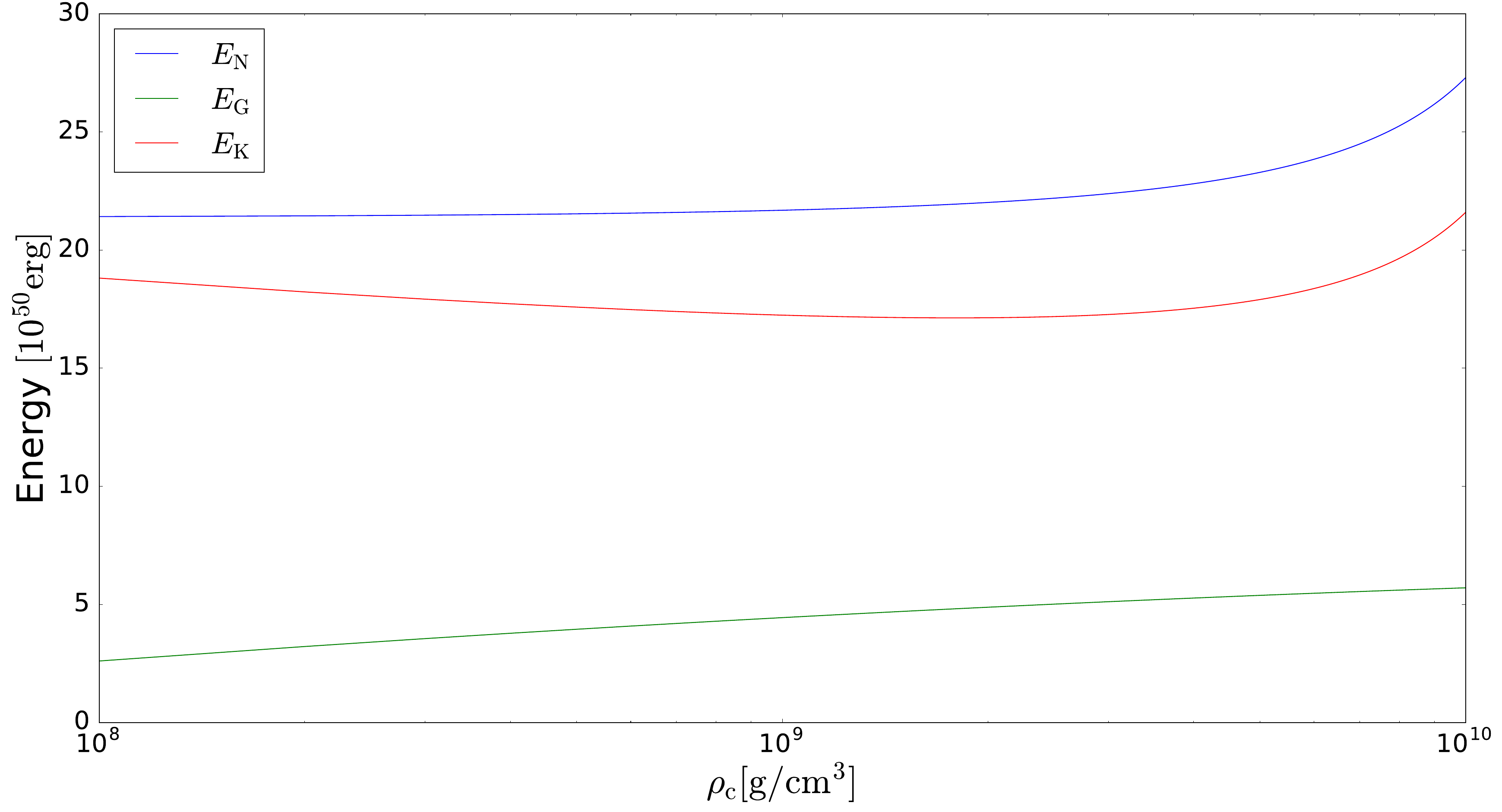}
\caption[]{Effect of varying $\rho_{c}$ on the nuclear, gravitational, and kinetic energies of a SN Ia. The other model parameters are held constant with values of: $M_{\mathrm{Ni}}\mathrm{=1.0}M_{\odot}$, $f_{\rm C/O}\mathrm{=0.0}$, $R_0\mathrm{=10^{9}~cm}$, $M_{\mathrm{ej}}\mathrm{=1.44}M_{\odot}$, and $\kappa \mathrm{=0.27 cm^2/g}$.}
\label{fig:EngrhoDep}
\vspace{-4ex}
\end{center}
\end{figure*}

There are strong constraints on the central density of the white dwarf progenitor, $\rho_{\mathrm{c}}$. The lower bound is due to there being a minimum density, for a given central temperature, at which the fusion of carbon can occur and trigger the thermonuclear explosion required for a SN Ia. For example, Fig.~1 of Ref.~\cite{Sahrling1994} calculates that even a high central temperature of $T_\mathrm{c}\mathrm{=4.0\times 10^8~K}$ requires a minimum central density around $\rho_{\mathrm{c}}\mathrm{=1.0\times 10^9~g/cm^{3}}$. The upper limit is provided by the point at which electron capture on nuclei to produce neutrons becomes a dominant process such that the white dwarf will collapse to a neutron star instead of becoming a SN Ia, which Refs.~\cite{Yoon2005} and \cite{Nomoto1991} calculate to occur around $\rho_{\mathrm{c}}\mathrm{\approx 1.0\times 10^{10}~g/cm^{3}}$. Figure~\ref{fig:EngrhoDep} shows that, over this constrained range, the kinetic energy of the system varies little with changes in $\rho_{\mathrm{c}}$. Therefore, it is not surprising that the light curves are only very weakly affected by changes in $\rho_{\mathrm{c}}$ as shown in the middle left panel of Fig. \ref{fig:param_dep}.

The fraction of unburned carbon/oxygen in the ejecta of SNe Ia is constrained observationally to be very low \citep{Thomas2011}, and therefore fractions above  5$\%$ will not be considered here. The lower left panel of Fig. \ref{fig:param_dep} shows that over this range, variation in $f_{\rm C/O}$ has little effect on the light curve of the SN Ia.

As varying the values of $f_{\rm C/O}$, $\rho_{\rm c}$, and $R_0$ within the physically motivated ranges mentioned above does not significantly affect the light curves, we will fix their values from here onwards at $f_{\rm C/O}=0.0$, $\rho_{\rm c}=1.0\times 10^9 \mathrm{g/cm^3}$, and $R_0=1.0\times 10^9 \mathrm{cm}$.

The upper right panel of Fig.~\ref{fig:param_dep} shows that, in general, increasing the total ejecta mass $M_{\mathrm{ej}}$ (while the mass of $^{56}\textup{Ni}$ is kept constant) decreases the brightness of the supernova and increases the light curve width. Specifically, when $M_{\mathrm{ej}}$ is doubled from $M_{\mathrm{ej}}=1.5M_{\odot}$ to $3.0M_{\odot}$ the peak luminosity of the SN Ia drops to about $\sim70\%$ of its former value, and the width of its light curve increases by $\sim60\%$. This behaviour is logical as a more massive ejecta would be more difficult for the radiation emitted in the decays of $^{56}\textup{Ni}$ and $^{56}\textup{Co}$ to pass through, so the same amount of energy escapes over a longer timescale,  mathematically represented in the equation for the light curve timescale:  $\tau_{\mathrm{m}} = \sqrt{2\kappa M_{\mathrm{ej}}/\beta c v_{\mathrm{sc}}}$. This results in a fainter, wider light curve. Note that the above expression treats the effect of $M_{ej}$ and $\kappa$ on $\tau_{\mathrm{m}}$ distinctly, so this physical interpretation is valid even though we held the effective opacity constant through the fixed value of $\kappa$. Variations in $M_{\mathrm{ej}}$ also affect the energetics of the supernova explosion, but the resulting effect on the light curve is negligible.

The middle right panel of Fig. \ref{fig:param_dep} shows that decreasing the mass of $^{56}\textup{Ni}$ from $M_{\mathrm{Ni}}\mathrm{=1.0}M_{\odot}$ to $M_{\mathrm{Ni}}\mathrm{=0.5}M_{\odot}$ in an $M_{{\rm ej}}=1.4M_{\odot}$ supernova causes the peak luminosity of the SN Ia to decrease to around half of its former value, while the timescale over which the SN Ia brightens and fades remains essentially unaffected. This is simply because the reduced amount of unstable $^{56}\textup{Ni}$ decreases the instantaneous power output from radioactive decay, making the SN Ia fainter at each point on the light curve. Variations in $M_{\mathrm{Ni}}$ also affect the energetics of the supernova explosion which determine the scale velocity $v_{\mathrm{sc}}$; however the resulting variation of $v_{\mathrm{sc}}$ is very small and has a negligible effect on the SN Ia light curve.

The lower right panel of Fig. \ref{fig:param_dep} shows that reducing the effective opacity of the SN Ia from $\kappa=0.30\ \mathrm{cm^2/g}$ to $\kappa=0.10\ \mathrm{cm^2/g}$ causes the width of the light curve to halve, while the peak luminosity nearly doubles. This reduction in $\kappa$ allows the radiation produced by the decay of $^{56}\textup{Ni}$ to escape more easily, which leads to the light curve peaking at an earlier time when a higher fraction of the initial $^{56}\textup{Ni}$ remains undecayed such that the instantaneous power from radioactive decay at the peak is higher and the SN Ia is therefore brighter. However, this also means that there is less trapped radiation still remaining at late times, such that the brightness of the SN Ia falls sharply soon after the peak.

\subsection{$M_{\rm Ni}$-$\kappa$ and width-luminosity relationships}
\label{ssec:MNikappa}

As mentioned at the end of Section \ref{ssec:LCM}, there is a relationship between the initial mass of \Ni in the ejecta $M_{\textup{Ni}}$ and the mean effective opacity of the ejecta $\kappa$ which allows us to eliminate $\kappa$ as an input parameter in our light curve model. To understand why this is the case, it is necessary to consider how the radioactive decay energy from the decay of $^{56}\textup{Ni}$ through to $^{56}\textup{\textup{Fe}}$ escapes the ejecta.

The radiation is initially produced as short-wavelength, high-energy gamma rays. The ejecta has a high opacity at these wavelengths, but a much lower opacity at longer wavelengths in the optical and infrared regions. Thus in order to escape, the radiation must somehow increase its wavelength. One process through which this increase in wavelength can happen is fluorescence, whereby an atom absorbs a single high energy, short wavelength photon and emits several lower energy, longer wavelength photons. This fluorescence process becomes less effective the more ionised a material becomes \cite{Pinto2000b}.

An increased $^{56}\textup{Ni}$ content increases the instantaneous power deposited into the ejecta by the radioactive decay of the $^{56}\textup{Ni}$. This increased power deposition results in the ejecta being heated more, and the higher temperature means the ejecta material will be more ionised, which in turn reduces the efficacy of the fluorescence process, so a larger fraction of the radiation remains trapped at short wavelengths.

In the light curve model, a larger fraction of the radiation remaining trapped can be expressed as an increased effective opacity $\kappa$. Thus there is a positive relationship between $M_{\textup{Ni}}$ and $\kappa$ where an increased $M_{\textup{Ni}}$ corresponds to an increased $\kappa$.

\begin{figure*}
\begin{center}
\includegraphics[width=\textwidth]{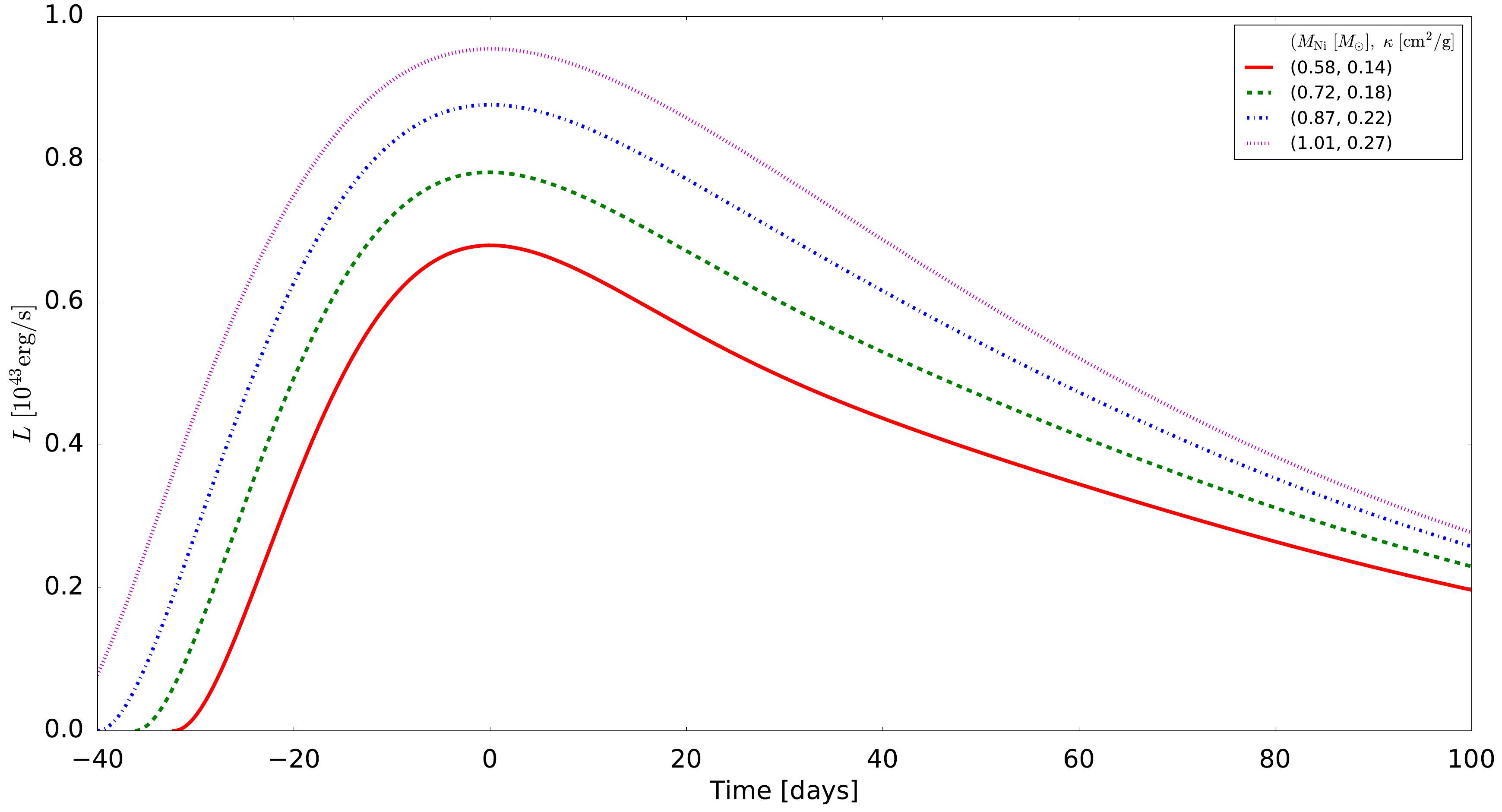}
\caption[]{Effect of varying both $M_{\mathrm{Ni}}$ and $\kappa$ simultaneously on the shape of the SN Ia light curve. The other model parameters are held constant with values of: $f_{\rm C/O}\mathrm{=0.0}$, $R_0\mathrm{=10^{9}~cm}$, $\rho_{\mathrm{c}}\mathrm{=1.0\times 10^9~g/cm^{3}}$, and $M_{\mathrm{ej}}\mathrm{=1.44}M_{\odot}$.}
\label{fig:MNikappaDep}
\vspace{-4ex}
\end{center}
\end{figure*}

As has been shown in Section \ref{ssec:LCparams} in the middle right and lower right panels of Fig. \ref{fig:param_dep}, increasing $M_{\textup{Ni}}$ in the light curve model increases the peak luminosity (i.e., height) of the light curve, and increasing $\kappa$ increases the timescale over which the SN Ia brightens and fades (i.e., the width of the light curve). Thus the above relationship between $M_{\textup{Ni}}$ and $\kappa$ means that an increase of these two parameters simultaneously increases both the height and width of the light curve, as can be seen in Fig.~\ref{fig:MNikappaDep}.

The above relationship between $M_{\textup{Ni}}$ and $\kappa$ can explain the WLR observed in SNe Ia \citep{Pinto2000b}. Since our model must be able to reproduce the WLR in order to be valid, this allows the specific relationship between $M_{\textup{Ni}}$ and $\kappa$ to be quantified.

\begin{figure*}
\begin{center}
\includegraphics[width=\textwidth]{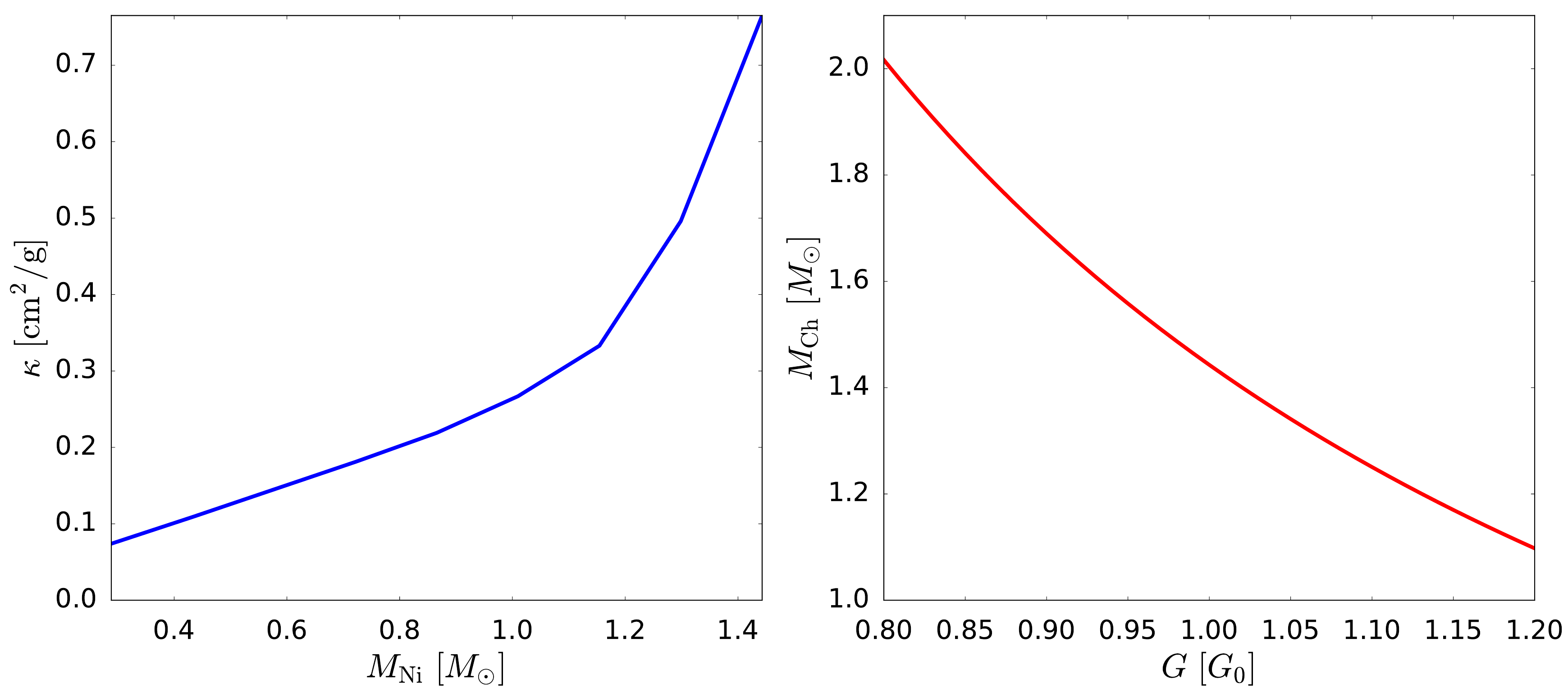}
\caption[]{{\it Left panel}: Relationship between $M_{\textup{Ni}}$ and $\kappa$ that is required to produce the observed WLR. {\it Right panel}: Relationship between Chandrasekhar mass $M_{\mathrm{Ch}}$ of a SN Ia expressed in units of solar mass $M_{\odot}\mathrm{=2.0\times 10^{30}~kg}$ and the local value of Newton's gravitational constant $G$, expressed in units of the standard value $G_0\mathrm{=6.67\times 10^{-11}N m^2 kg^{-2}}$. The standard Chandrasekhar mass value is given by $M_{\mathrm{Ch}}(G_0)\mathrm{=1.44}M_{\odot}$.}
\label{fig:MNivkappa_MChvG}
\vspace{-4ex}
\end{center}
\end{figure*}

The exact $M_{\textup{Ni}}$-$\kappa$ relationship that is required to produce the observed WLR can be found by calculating, for a given increase in $M_{\textup{Ni}}$, how much of an increase in $\kappa$ is required to produce a new curve whose peak luminosity and width are both scaled from the peak luminosity and width of the old curve by the same factor. This can then be repeated, such that the required $\kappa$ values can be calculated for many $M_{\textup{Ni}}$ values, therefore establishing a relationship between $M_{\textup{Ni}}$ and $\kappa$. In practice this is done as follows: consider a light curve A that is produced with parameters ${M_{\textup{Ni}}}_{\mathrm{A}}$ and $\kappa_{\mathrm{A}}$, and a set of light curves $\rm B_i$ produced with fixed ${M_{\textup{Ni}}}_{\mathrm{B}}$ but different $\kappa_{\mathrm{B, i}}$. The accepted $\kappa_{\mathrm{B, i}}$ is the one that produces the light curve $\rm B_i$ whose shape about the peak most closely matches that of light curve A (computed by minimising $\chi^2$) once light curve $\rm B_i$ has been stretched in both height and width by a factor $s_{\rm B,i}$. This process yields the new pair of values (${M_{\textup{Ni}}}_{\mathrm{B}}$, $\kappa_{\mathrm{B, accepted}}$) and can be repeated for other values ${M_{\textup{Ni}}}_{\mathrm{C}}$ and so on in order to build a relationship between $M_{\textup{Ni}}$ and $\kappa$. It does require a single point in the $M_{\textup{Ni}}$-$\kappa$ parameter space, $({M_{\textup{Ni}}}_{\mathrm{A}}, \kappa_{\mathrm{A}})$, through which the relationship passes, to be specified. Following Refs.~\cite{Piroetal2009} and \cite{Childressetal2015}, we specify the point $(M_{\mathrm{Ni}}, \kappa)=(0.79\ M_{\odot}, 0.20\ \mathrm{cm^2/g})$. The resulting relationship between $M_{\textup{Ni}}$ and $\kappa$ is displayed in the left panel of Fig. \ref{fig:MNivkappa_MChvG}.

Using this calculated $M_{\textup{Ni}}$-$\kappa$ relationship, we can reproduce a set of light curves for a population of SNe Ia in a standard gravity environment with varying $M_{\textup{Ni}}$ values. We can then verify that this set of light curves does obey the WLR by rescaling them. The rescaling is done by stretching the light curve's height and width by a factor $s$ such that the shape of the stretched light curve about the peak most closely matches that of a template light curve (again computed by minimising $\chi^2$). As confirmed by the tight distribution of peak luminosities in the rescaled light curves seen in the right panel of Fig.~\ref{fig:G0_Rescaling}, which shows the unscaled and rescaled light curves for a population of SNe Ia in standard gravity, the WLR is obeyed for $G=G_0$ when the $M_{\textup{Ni}}$-$\kappa$ relationship calculated as described above is used. This is expected, given the way in which the $M_{\textup{Ni}}$-$\kappa$ relationship was calculated. Nonetheless, checking that the population does obey the WLR confirms the $M_{\textup{Ni}}$-$\kappa$ relationship has been calculated correctly and gives a clear demonstration of what the $M_{\textup{Ni}}$-$\kappa$ relationship achieves. If the value of $G$ is constant (i.e. $G(z)=G_0$) then the SNe Ia will have the same average rescaled intrinsic peak luminosity at all redshifts, and therefore act as conventional standardisable candles.

\subsection{Gravitational dependence of light curve}
\label{ssec:GdepLCM}

Now that a model for producing the light curve of a SN Ia for a given set of input parameters has been established, the next step is to understand how the values of these input parameters depend on the local strength of gravity. As an initial investigation, this paper focuses mainly on the dependence of the total ejecta mass $M_{\mathrm{ej}}$ on the strength of gravity. This is the most obvious and important effect the strength of gravity should have on any of the input parameters. A simple measure of the strength of gravity is through the value of Newton's gravitational constant, $G$, with a smaller value of $G$ corresponding to a weaker gravity and vice versa.

As mentioned previously, the SN Ia occurs when the white dwarf progenitor accretes enough mass from a binary partner to increase the core temperature and density above the level at which carbon fusion can occur. The underlying physics results from the white dwarf being made of degenerate matter that obeys an inverse mass-radius relationship such that the additional mass has the effect of decreasing the white dwarf's radius, thereby increasing the density and temperature in its core. The consensus model for SNe Ia is that the critical mass at which the carbon detonation is triggered, and therefore $M_{\mathrm{ej}}$, is approximately equal to the mass at which the internal electron degeneracy pressure that prevents the white dwarf from collapsing under its own weight can no longer withstand the inwards force of gravity, a mass known as the Chandrasekhar mass, $M_{\mathrm{Ch}}$ \citep{Hillebrandt2000}. Making the assumption $M_{ej} \approx M_{\mathrm{Ch}}$, it becomes necessary to understand how $M_{\mathrm{Ch}}$ depends on $G$.

$M_{\mathrm{Ch}}$ can be calculated by equating the inwards force of gravity against the outwards force due to the white dwarf's internal electron degeneracy pressure \citep{Chandrasekhar1931}. The electron degeneracy pressure arises from the Pauli exclusion principle that states no two fermions can occupy the same quantum mechanical state. Therefore, when several electrons are confined to a small volume they must each occupy different energy levels, and adding further electrons to this small volume by compressing the material raises the energy of the highest occupied level. This means that energy is required to compress the electrons, which is the definition of a pressure, in this case known as electron degeneracy pressure. A condensed derivation of the equation for the Chandrasekhar mass is presented here, but a fully detailed version can be seen in Appendix \ref{App:MChDerivation}. The derivation of the equation for the Chandrasekhar mass is very well known, for example see \cite{MChDerivation}.

The equation of hydrostatic equilibrium for a spherically symmetrical stellar fluid in Newtonian gravity can be written as
\begin{equation}
\frac{1}{r^{2}} \frac{{\rm d}}{{\rm d}r} \left( \frac{r^{2}}{\rho} \frac{{\rm d}P}{{\rm d}r} \right) = - 4\pi{G}\rho,
\label{eq:stellarstate1}
\end{equation}
where $r$ is the radial coordinate, $P$ is the pressure of the fluid, $\rho$ is the density of the fluid. For the degenerate material under high amounts of compression inside a white dwarf, the electrons will have a large energy due to the electron degeneracy pressure, and so will have a velocity approaching that of the speed of light. Thus, the white dwarf material is best described as a relativistic Fermi gas with an equation of state given by
\begin{equation}
P = \frac{\hbar c}{12 \pi^{2}} \left( \frac{3 \pi^{2} \rho}{m_{\mathrm{N}} \mu} \right)^{4/3} \equiv K\rho^{4/3},
\label{eq:EoS1}
\end{equation}
where $\hbar$ is the reduced Planck constant, $m_{\mathrm{N}}$ is the nucleon mass, and $\mu=\left\langle A/Z \right\rangle$ is the average mass number per nuclear charge with $\mu \approx2$ for the $^{12}\mathrm{C}$ and $^{16}\mathrm{O}$ that make up the majority of the white dwarf. This equation of state is of the form of a polytrope $P = K \rho^{\gamma}$ with $\gamma=4/3$. By defining $\rho \equiv \lambda \Theta^n$, $\gamma \equiv \frac{n+1}{n}$, and introducing a radial variable $y\equiv r/\alpha$ where $\alpha \equiv \sqrt{(n+1)K \lambda^{(1-n)/n}/4 \pi G}$, Eq.~(\ref{eq:stellarstate1}) becomes the Lane-Emden equation for polytropes in hydrostatic equilibrium \citep{Lane1870}:
\begin{equation}
\frac{1}{y^2} \frac{{\rm d}}{{\rm d}y} \left(y^2 \frac{{\rm d}\Theta}{{\rm d}y} \right) = - \Theta^n~.
\label{eq:LaneEm1}
\end{equation}
The white dwarf mass is given by
\begin{equation}
M = 4 \pi \lambda \alpha^3 \left[-y^2 \frac{{\rm d}\Theta}{{\rm d}y} \right]_{y_1},
\label{eq:Mint11}
\end{equation}
where $y_1$ corresponds to the outer radius of the star $R$ where $\rho(R)=\Theta(y_1)=0$, and can be found by numerically solving Eq.~(\ref{eq:LaneEm1}), which gives $y_1(n=3)=6.89685$ and $-y^2{\rm d}\Theta/{\rm d}y\mid_{y_1}=2.01824$ for a white dwarf. Substituting the definitions for $\lambda$ and $\alpha$ into Eq.~(\ref{eq:Mint11}) leads to
\begin{equation}
M_{\mathrm{Ch}}=\frac{\sqrt{3 \pi}}{2} {\left( \frac{\hbar c}{G} \right)}^{3/2} \frac{1}{( \mu m_{\mathrm{N}} )^2} \left[- y^2 \frac{{\rm d}\theta}{{\rm d}y} \right]_{y_1},
\label{MCh1}
\end{equation}
which defines the Chandrasekhar mass $M_{\mathrm{Ch}}$. It can be seen that there is a clear proportionality for the Chandrasekhar mass, and therefore in our model, the white dwarf mass at supernova: $M_{\mathrm{ej}}=M_{\mathrm{Ch}}\propto G^{-3/2}$, {as was shown in earlier works \cite{Amendolaetal1999,Garcia-Berroetal1999,RiazueloUzan2002}}. This result is displayed graphically in the right panel of Fig. \ref{fig:MNivkappa_MChvG}. The fact that the Chandrasekhar mass $M_{\rm Ch}$ increases with a decreasing $G$ can be understood as follows: when $G$ decreases, gravity per unit mass becomes weaker, and therefore the electron degeneracy pressure can counteract against the gravity produced by more mass before the collapse occurs.

In addition, we note that the relationship $y \equiv r/\alpha = \mathrm{const}$ where $\alpha \equiv \sqrt{(n+1)K \lambda^{(1-n)/n}/4 \pi G}$ tells us that the radius of the white dwarf $R$ also depends on $G$ as $R \propto G^{-1/2}$. However, since changes to the white dwarf radius, and therefore the model parameter $R_0$, have a negligible effect on the light curve, this effect can be neglected.

Moreover, the binding energy $E_G$ used in calculating the energetics of the SN Ia explosion is also affected by changes in the strength of gravity. We use a simplified model that assumes constant density of the white dwarf, but a more complicated model is not necessary as the effect is at the 1-2$\%$ level. The binding energy is related to the strength of gravity, the mass of the progenitor, and the radius of the progenitor by
\begin{align}
E_G \propto \frac{GM^2}{R}.
\end{align}
The above results give $M \propto G^{-3/2}$ and $R \propto G^{-1/2}$, and they lead to $E_G \propto G^{-3/2}$. This dependence of binding energy on the strength of gravity affects the light curve in non-standard $G$ through the kinematics of the explosion, but as mentioned above this effect is minor in comparison to the dominant effect of the gravitational-dependence of the Chandrasekhar mass.

It should be noted that a varying $G(z)$ would normally affect the background cosmological evolution through the scale factor $a$, and this would change the best fit cosmological parameters computed with the theoretical distance-redshift relation, potentially having a strong effect on structure formation as well. However, in this work we shall not consider the effect of varying $G$ on the background expansion rate, as there are theories in which the variation of $G$ mainly affects interactions of matter particles, for example in the form of a Yukawa-type fifth force which decays at large distance, such as viable $f(R)$ gravity models \cite{Braxetal2008,Wangetal2012,Ceron-Hurtadoetal2016}
that feature a working chameleon screening mechanism \cite{KhouryandWeltman20041,KhouryandWeltman20042}. Also, when considering the time variation of $G$, we assume that $G(z)$ only varies noticeably on cosmological time scales -- the supernova timescale is much shorter, over which we take $G$ as constant. Note that in more general modified gravity models the internal structure of the star can be affected as well, see for example \cite{Babichevetal2016}, but this possibility is beyond the scope of this initial study.

Previous works \cite{Garcia-Berroetal1999,RiazueloUzan2002} made the assumption $M_{\rm Ni} \propto M_{\rm Ch}$ and therefore $M_{\rm Ni} \propto G^{-3/2}$. However, here we do not assume any systematic dependence of $M_{\rm Ni}$ on $M_{\rm Ch}$ or $G$. Hence when we compare populations of SNe Ia in different strengths of gravity, as in the subsequent section, we do not change the set of $M_{\rm Ni}$ values used to generate the varied members of the population for each new value of $G$. A proper investigation into how the amount of $^{56}{\rm Ni}$ produced would differ when $G \neq G_0$ would require a detailed study of star formation and full hydrodynamical modeling of the SN Ia explosion in such scenarios, and if the evolution of $G$ over the relevant timescales is non-negligible then the situation is even more complicated -- such an investigation is far beyond the scope of this work.

As we mainly want to consider the effect of the gravitational dependence of $M_{\rm ej}$ on the WLR, we make the assumption that the $M_{\textup{Ni}}$-$\kappa$ relationship that underlies the observed WLR is a consequence of nuclear and atomic physics, and does not depend on the local strength of gravity around the SN Ia. Considering the astrophysics discussed in Section \ref{ssec:MNikappa}, this assumption is equivalent to assuming that changes in the strength of gravity around the SN Ia do not significantly affect the transport of radiation through the ejecta, which we believe is a reasonable assumption to make for this initial investigation.

Thus with the dominant Chandrasekhar mass effect and the minor binding energy effect we have defined a gravity-dependent light curve model that we can use to investigate the effects of gravity on the interpretation of SN Ia cosmology data. We will do this in the next section.

\section{Reinterpreting SN Ia Cosmology}\label{sec:SN Iacosmo}

SN Ia cosmology relies heavily on the rescaling procedures that allow SNe Ia to act as standardisable candles. Now that we have a gravity-dependent light curve model, we are able to investigate whether the WLR and our standardisation procedure are affected in non-standard gravity where $G\neq G_0$.

\subsection{Gravitational effect on WLR and standardisability}\label{ssec:GDepWLR}

\begin{figure*}
\begin{center}
\includegraphics[width=\textwidth]{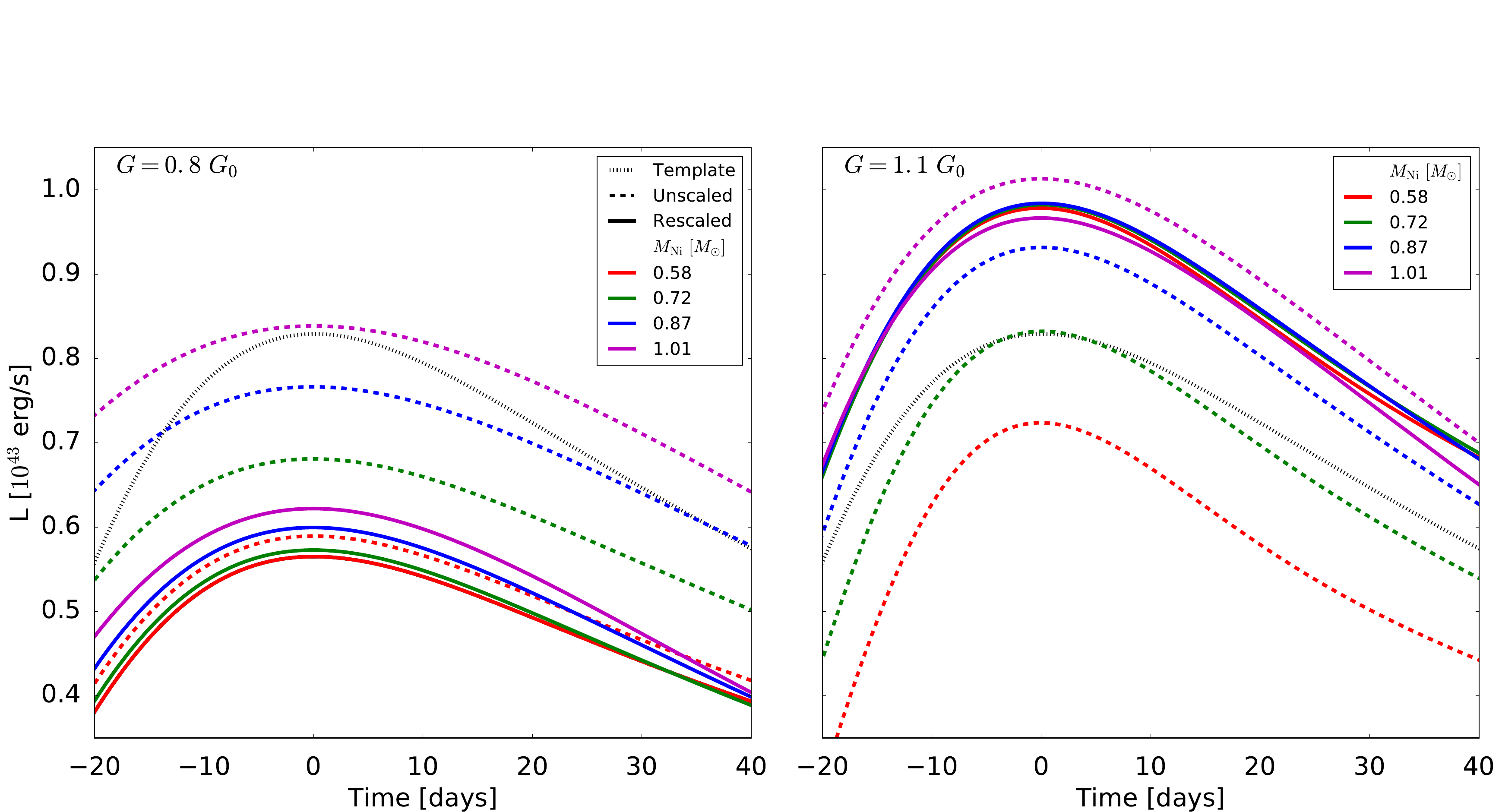}
\caption[]{Rescaling of light curves from a SN Ia population with varying $M_{\mathrm{Ni}}$ in non-standard gravity ($G\neq G_0$). A template curve, whose shape around the peak the other curves have been rescaled to match with, is also shown. {\it Left panel}: $G=0.8G_0$. {\it Right panel}: $G=1.1G_0$.}
\label{fig:G_Rescaling}
\vspace{-3ex}
\end{center}
\end{figure*}

Rescaling a population of SNe Ia in a non-standard gravity environment generated using the same set of $M_{\rm Ni}$ values and the same calculated $M_{\textup{Ni}}$-$\kappa$ relationship as used for the standard gravity SNe Ia still produces a set of approximately uniform rescaled curves, as can be seen in Fig. \ref{fig:G_Rescaling} for two different values of $G$: a weaker gravity ($G=0.8G_0$; left panel) and a stronger one ($G=1.1G_0$; right panel). Therefore, the WLR is reproduced even for $G \neq G_0$.

Interestingly, Fig.~\ref{fig:G_Rescaling} shows that the average rescaled intrinsic peak luminosity of the non-standard gravity SN Ia population is different from that of the standard gravity population as seen in the right panel of Fig.~\ref{fig:G0_Rescaling}. The black dotted lines in Fig.~\ref{fig:G_Rescaling} are the same template curves as used in the right panel of Fig.~\ref{fig:G0_Rescaling}, thus we see that a weaker (stronger) gravity leads to intrinsically less (more) luminous SNe Ia after applying the same shape matching rescaling procedure.

\begin{figure*}
\begin{center}
\includegraphics[width=\textwidth]{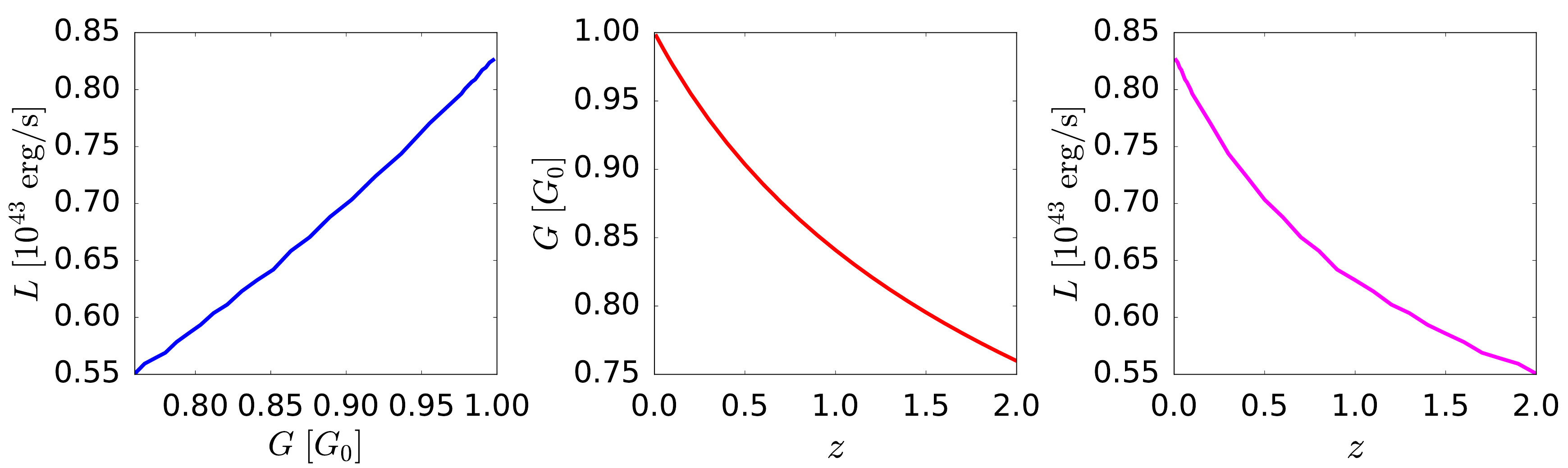}
\caption[]{{\it Left panel}: The $L(G)$ relationship showing the effect of the local strength of gravity on the average rescaled intrinsic peak luminosity of a SN Ia population. {\it Centre panel}: The $G(z)$ relationship for our numerical example specified by Eq.~(\ref{eq:Gvz}) with $n=4$. {\it Right panel}: The $L(z)$ relationship that results from using our gravity-dependent light curve model to investigate how the average rescaled intrinsic peak luminosity $L$ varies with redshift in a model of gravity that varies as specified by Eq.~(\ref{eq:Gvz}) with $n=4$.}
\vspace{-3ex}
\label{fig:LGz}
\end{center}
\end{figure*}

Therefore, our gravity-dependent light curve model allows us to obtain a relationship between the average rescaled intrinsic peak luminosity and the strength of gravity, which is displayed in the left panel of Fig.~\ref{fig:LGz}. This means that in a model where the local strength of gravity varies as a function of redshift (i.e. $G(z) \neq G_0$) the average rescaled intrinsic peak luminosity will be different at different redshifts, and thus the SNe Ia will no longer be conventional standardisable candles. Following the arguments made in Section \ref{sec:intro}, interpreting the SN Ia cosmology observations in the context of such a model will lead to different SN Ia luminosity distances from the values obtained by assuming $G$ is independent of redshift (i.e. $G(z) = G_0$), and therefore a different best-fitting cosmology.

This is the key result of this work, and in the next subsection, as an example of its use, we will apply it to a toy model where the value of $G$ changes over time to see what the consequences would be on the distance-redshift relation that is calculated from the existing SN Ia data.

Previous studies of the effects of varying G on SN Ia luminosity assumed either that there was some unknown positive relationship between the peak luminosity of any individual SN Ia $\mathcal{L}_{\rm peak}$ and the Chandrasekhar mass such that $\mathcal{L}_{\rm peak} = {M_{\rm Ch}}^{\gamma} \propto G^{-3\gamma/2}$ where $\gamma > 0$ \cite{Amendolaetal1999} or, as mentioned in Section~\ref{ssec:GdepLCM}, assumed that the amount of $^{56}$Ni produced in any individual SN Ia (which determines the peak luminosity) is directly proportional to the Chandrasekhar mass such that $\mathcal{L}_{\rm peak} \propto M_{\rm Ni} \propto M_{\rm Ch} \propto G^{-3/2}$ \cite{Garcia-Berroetal1999,RiazueloUzan2002} with the result that both methods suggest $\mathcal{L}_{\rm peak}$ will increase as $G$ decreases. Refs.~\cite{Garcia-Berroetal1999,RiazueloUzan2002} also derived a simple proportionality between $G$ and the width of the light curves ($\tau$ in their work), although this was largely irrelevant to their analysis as the works are simplified phenomenological studies where the impact of varying $G$ on SN Ia standardisation procedures (which typically involve the width of the light curve) was not taken into account.

In contrast, we do not derive a simple proportionality between the intrinsic peak luminosity and $G$, but instead aim to bring the analysis of SN Ia cosmology in modified gravity closer to observational work by using a simplified standardisation procedure on a set of SNe Ia light curves in environments with different values of $G$ to investigate empirically how the average of the intrinsic peak luminosities of the standardised light curves (the quantity that is essentially used in SN Ia cosmology) is affected by varying $G$.
Note that because our simplified standardisation procedure is based on the WLR, the underlying physics behind which is the $M_{\rm Ni}-\kappa$ relationship discussed in Section~\ref{ssec:MNikappa}, any change in $M_{\rm Ni}$ due to varying $G$ will not significantly affect the standardised peak luminosity as a light curve produced with a different $M_{\rm Ni}$ will always closely match the $G=G_0$ template after rescaling due to the $M_{\rm Ni}-\kappa$ relationship. Instead the dominant effect varying $G$ has on the rescaled peak luminosities is through $M_{\rm Ch}$. In Section~\ref{ssec:GdepLCM} we showed that decreasing $G$ will increase $M_{\rm Ch}$, and the top right panel in Fig.~\ref{fig:param_dep} shows that this increases the width of the light curve but decreases the luminosity compared to the $G=G_0$ template. Therefore during standardisation when the width of the $G<G_0$ curve is decreased to match the shape of the $G=G_0$ template around its peak, the equal decrease in the luminosity pushes the rescaled peak luminosity even further below that of the template.
This intuitively explains why our analysis concludes that decreasing $G$ leads to a decrease in the average rescaled intrinsic peak luminosity $L$ and vice versa.

As previously mentioned, the standardisation procedures used in observational SN Ia cosmology are more complex than our rescaling method. Typically such studies describe the variability of SNe Ia as being captured by two parameters: the time-stretching of the light curve and the colour of the SN Ia at peak brightness \cite{MLCS,SALT2}. Note that we cannot consider SN Ia colour in our simplified procedure because the semi-analytical model we use can only produce UVOIR light curves. The dependence of the absolute magnitude on the two parameters can be estimated using an entire sample of SNe Ia simultaneously, not just low-redshift SNe Ia whose distances can be computed from other methods. Any remaining dispersion in the absolute magnitudes of a sample after this stretch-color standardisation could be used in combination with our model to constrain the variation of $G$ more accurately than has been done in previous works \cite{Gaztanagaetal2002,LorenAguilaretal2003,MouldUddin2014}.

\subsection{Numerical example}
\label{ssec:NumEx}

For our numerical example we specify a $G(z)$ relationship as displayed in the centre panel of Fig.~\ref{fig:LGz} where the value of $G$ changes as
\begin{align}\label{eq:Gvz}
G(z) = G_0 (1+z)^{-1/n},
\end{align}
with $n=4$. Note that this is a toy model intended to describe the late-time variation of $G$ and we do not assume that it works for $z>2$-$3$. We then use our gravity-dependent light curve model to compute the corresponding variation in the average rescaled intrinsic peak luminosity $L$ with redshift that occurs as a consequence of our non-constant $G$, and this is shown in the right panel of Fig.~\ref{fig:LGz}.

\begin{figure*}
\begin{center}
\includegraphics[width=\textwidth]{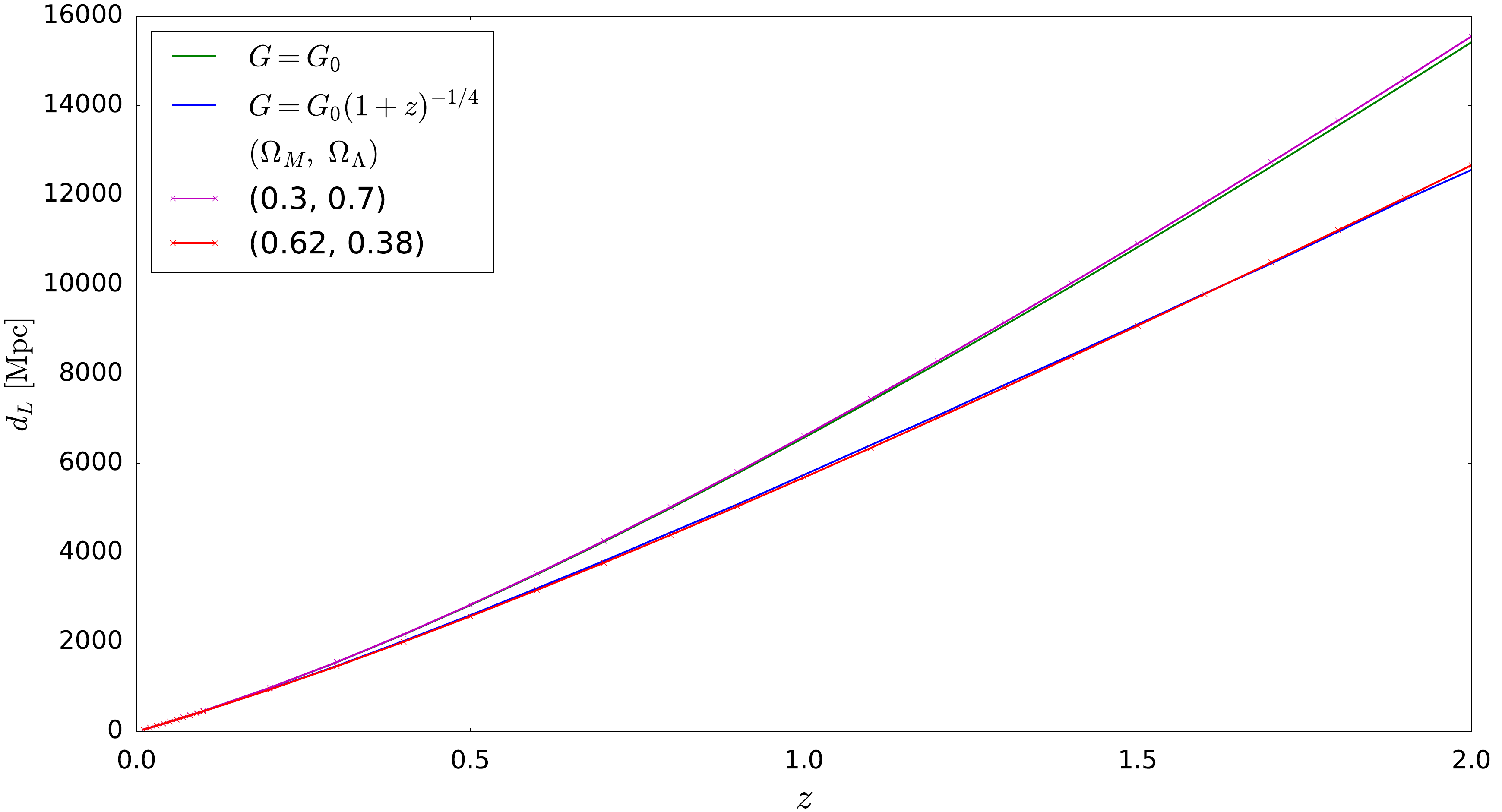}
\caption[]{The $d_L(z)$ relationship showing that when the SN Ia apparent magnitude-redshift data is (re)interpreted for a universe where $G(z)=G_0(1+z)^{-1/4}$ the new best fitting cosmology is $(\Omega_{\rm M}, \Omega_{\Lambda})=(0.62, 0.38)$ when the Universe is assumed to be flat. The line labelled as `$G=G_0$' is computed assuming a standard cosmology of $(\Omega_{\rm M}, \Omega_{\Lambda})=(0.31, 0.69)$ to distinguish from the purple line and to compare the effect of slightly varying $\Omega_{\rm M}$ with that of reinterpreting the data assuming $G(z)=G_0(1+z)^{-1/4}$.} \label{fig:dvz}
\end{center}
\end{figure*}

Once we have this $L(z)$ relationship we can convert the intrinsic luminosity values to values of absolute magnitude $M$ with
\begin{equation}
\label{eq:lumabsmag1}
M = M_{\odot} - 2.5\log_{10} \left( \frac{L}{L_{\odot}} \right),
\end{equation}
and use these absolute magnitude values to (re)interpret the existing SN Ia apparent magnitude-redshift data $m(z)$ and produce the distance-redshift relation using
\begin{equation}\label{eq:dLobs}
d_L = 10^{(m-M-25)/5}.
\end{equation}
The distance-redshift relationship that results from (re)interpreting the existing SN Ia data in a model where gravity varies with redshift as $G(z) = G_0 (1+z)^{-1/4}$ is shown in Fig.~\ref{fig:dvz}. This re-interpreted $d_L(z)$ can then be compared to theoretical distance-redshift relationships produced using
\begin{widetext}
\begin{equation}\label{eq:d_Lz}
d_{\mathrm{L}}(z;\Omega_{\mathrm{M}},\Omega_{\mathrm{\Lambda}}, H_0) = \frac{c(1+z)}{H_0 \sqrt{\left|\kappa\right|}}
S \left[ \sqrt{\left|\kappa\right|} \int^{z}_{0} \left[ (1+z^{\prime})^2 (1+\Omega_{\mathrm{M}} z^{\prime}) - z^{\prime}(2+z^{\prime})\Omega_{\mathrm{\Lambda}} \right]^{-1/2} {\rm d}z^{\prime} \right],
\end{equation}
\end{widetext}
where the function $S(x)$ is given by
\begin{equation}
S(x) =
\begin{cases}
\sin (x) & \text{if } \Omega_{\mathrm{M}} + \Omega_{\mathrm{\Lambda}} > 1,
\\
\sinh (x) & \text{if } \Omega_{\mathrm{M}} + \Omega_{\mathrm{\Lambda}} < 1,
\\
x & \text{if } \Omega_{\mathrm{M}} + \Omega_{\mathrm{\Lambda}} = 1,
\end{cases}
\end{equation}
and the curvature $\kappa$ is
\begin{equation}
\kappa =
\begin{cases}
1 & \text{if } \Omega_{\mathrm{M}} + \Omega_{\mathrm{\Lambda}} = 1,
\\
1 - \Omega_{\mathrm{M}} - \Omega_{\mathrm{\Lambda}} & \text{otherwise},
\end{cases}
\end{equation}
where $\Omega_{\mathrm{M}}$ and $\Omega_{\mathrm{\Lambda}}$ are the matter and dark energy density parameters, and $H_0$ is the current day (i.e. at $z=0$) Hubble parameter value.

In Fig.~\ref{fig:dvz} we start by producing a distance-redshift relation for a standard $\Lambda$CDM cosmology with $(\Omega_{\rm M}, \Omega_{\Lambda})=(0.31, 0.69)$ (green line) produced using Eq.~(\ref{eq:d_Lz}) and construct a mock apparent magnitude-redshift relation using Eq.~(\ref{eq:dLobs}) assuming $G(z)=G_0$ and therefore a redshift-independent $M$. Once we have this mock data we can compute the $d_L(z)$ relation required to match it if we assume that $G(z)=G_0(1+z)^{-1/4}$ which means that intrinsic peak luminosity $L$, and therefore the absolute magnitude at peak $M$ through Eq.~(\ref{eq:lumabsmag1}), depends on redshift (blue line). This gravitational effect means that the supernovae at higher redshift are intrinsically fainter, which means that less cosmic acceleration is required to explain the dimming of distant SNe Ia. We then use a curve fitting approach to identify the values of the cosmological parameters $(\Omega_{\rm M}, \Omega_{\Lambda})$ in Eq.~(\ref{eq:d_Lz}) that give the best fit to the reinterpreted mock data. We find that the new best fitting cosmology is $(\Omega_{\rm M}, \Omega_{\Lambda})=(0.62, 0.38)$ when the Universe is assumed to be flat (red line). We also display the distance-redshift relation produced using Eq.~(\ref{eq:d_Lz}) with $(\Omega_{\rm M}, \Omega_{\Lambda})=(0.3, 0.7)$ (purple line) to compare the effect of slightly varying $\Omega_{\rm M}$ with that of reinterpreting the data for a redshift-dependent $G$.

This is just a numerical example to demonstrate the importance of properly considering the effects of modified gravity on supernova cosmology in models with a specified variation of the strength of gravity with redshift $G(z)$.

The process could also be reversed to identify what $G(z)$ relationship would be required for the existing SN Ia data to infer a given cosmology, for example one with $(\Omega_{\rm M}, \Omega_{\Lambda})=(1.0, 0.0)$ where there would be no sign of an accelerating Universe and the dimming of distant supernova is purely intrinsic. We will investigate this possibility in a separate work.

\section{Conclusion}
\label{sec:conc}

Modified gravity theories have been studied as interesting alternatives to the standard $\Lambda$CDM paradigm to explain the late-time cosmic acceleration, the first evidence for which resulted from analyses of the dimming of distant Type Ia supernovae (SNe Ia). In most such studies, the modification to Einsteinian gravity is assumed to provide a mechanism to accelerate the expansion rate of the Universe, without affecting the interpretation of the Type Ia supernova (SN Ia) data itself. However, given the diversity of modified gravity theories being studied in the literature, it is not unnatural to envisage that at least some of these models may affect the properties of the progenitors of SNe Ia -- Chandrasekhar-mass white dwarfs, where gravity plays an important role -- and therefore the properties of the SNe Ia themselves. That being the case, the interpretation of SN Ia data would be affected, which can have nontrivial consequences in cosmology.

In this paper, we make an updated attempt to understand the effect of a non-standard gravity on SNe Ia. Because modern supernova cosmology depends on the ability of SNe Ia to behave as standardisable candles, we are primarily interested in the standardisability of SNe Ia light curves in both standard and non-standard scenarios of gravity. Previous work \cite{Amendolaetal1999,Garcia-Berroetal1999,RiazueloUzan2002} suggested a straightforward proportionality between the intrinsic luminosity of SNe Ia and the value of Newton's constant $G$. We advance this method by investigating how the full SNe Ia light curves are affected by modified gravity, an approach that allows us to test whether the width-luminosity relation (WLR) is reproduced, and verify that the standardisation procedure, vital for the use of SNe Ia as measures of distance, still works when $G \neq G_0$. As a first step towards a more comprehensive study, this work has made three simplifications that ideally should be revisited in more detail in future investigations. First, we use a semi-analytical model to predict the SNe Ia light curves, although a more detailed analysis may involve running full 3D hydrodynamical simulations for the SN Ia explosion. The physics and astrophysics of SNe Ia is not yet fully understood, but the simple light curve model used here works reasonably well empirically. Second, we model the effect of modified gravity as a time variation of the gravitational constant $G$, which affects properties of the white dwarf progenitors such as their mass, radius and gravitational binding energy. In principle, a proper treatment of a given modified gravity model requires us to solve the equation for hydrostatic equilibrium, which may lead to changes of internal structures of the white dwarf. In our treatment, however, the main effect is a change of the Chandrasekhar mass as $M_{\rm Ch}\propto G^{-3/2}$. We note that a time variation of $G$ is a prediction of many modified gravity theories and indeed a key feature of early theoretical models such as the Brans-Dicke theory. Lastly, we use a light curve standardisation procedure that is simplified in comparison to those that are used in observational SN Ia cosmology. Our procedure involves rescaling light curves so that the shape about their peaks matches that of a template light curve. This simplification is only used for this proof-of-concept study, and for a more rigorous analysis it is necessary to closely follow the procedure used in real observations. We leave this possibility for future work.

The semi-analytical light curve model described in Eq.~(\ref{eq:luminosity31}) depends on a few physical parameters, including the ejecta mass $M_{\rm ej}$, the nickel-56 mass $M_{\rm Ni}$, the effective opacity of the ejecta $\kappa$, the progenitor radius $R_0$, the central density of the ejecta $\rho_{\rm c}$ and the mass fraction of unburned carbon-oxygen $f_{\rm C/O}$. Fortunately, the last three have little impact on the light curve in their constrained range of values and, in the case of $R_0$, in the range of variations of its value as caused by varying $G$. In addition, in this model $\kappa$ is both physically expected to strongly correlate with $M_{\rm Ni}$ and is required to do so in order to explain the width-luminosity relation of SNe Ia, which is fundamental to the standardisability of their light curves. We have derived such a $M_{\rm Ni}$-$\kappa$ relation and verified that it allows us to standardise the SNe Ia light curves (i.e. match both the shape and amplitude of the light curves) for a range of $M_{\rm Ni}$ values when the Chandrasekhar mass is fixed to its standard value of $1.44M_\odot$.

However, if $G$ deviates from its present-day value $G_0$, then the Chandrasekhar mass differs from $1.44M_\odot$ as well so that the supernova ejecta will have a different $M_{\rm ej}$. To first order, we assume that the $M_{\rm Ni}$-$\kappa$ relation established above (or observationally from nearby SNe Ia, for which $G\approx G_0$) is still valid because it is defined by physics involving interactions other than gravity. In this case, we find that rescaling the SNe Ia light curves by matching their shapes about the peak when $G \neq G_0$ still leads to good match of their heights (with a slightly larger spread than for $G=G_0$ in the latter), thus verifying that the WLR holds for $G \neq G_0$. However, these rescaled peak luminosities are different from the `template' value (which is obtained by using nearby SNe Ia and therefore for $G=G_0$). In other words, a time variation of $G$ results in a time dependence of the average rescaled intrinsic peak luminosity of SNe Ia, so that they are no longer standardisable candles in such a scenario.

Because of this, in a theory where $G$ is time-dependent, the observed dimming of distant SNe Ia will at least partly be due to the variation of the local strength of gravity, and is not entirely the consequence of the cosmic acceleration. We have demonstrated this by using two models, one of standard $\Lambda$CDM with $(\Omega_{\rm M}, \Omega_{\Lambda})=(0.3, 0.7)$ and no time variation of $G$, and a `modified' $\Lambda$CDM with $(\Omega_{\rm M}, \Omega_{\Lambda})=(0.62, 0.38)$ and $G(z)=G_0(1+z)^{-1/4}$. Both scenarios predict the same apparent magnitude-redshift relation for SNe Ia, and therefore cannot be distinguished using supernova cosmology. This means that for accurate cosmological tests of gravity we must carefully consider gravity's impact on SN Ia astrophysics, and how this consequently loops back to affect the interpretations of the cosmological data being used. This may have an impact on potential resolutions of the recent observational tensions in the concordance $\Lambda$CDM model, a possibility that we will study in future works.

This also brings us to another interesting question: can a cosmological model with a time-varying $G$ but no $\Lambda$ fit SN Ia data? As {computed} in this paper, since a weak gravity leads to an intrinsically fainter SN Ia population{, if $G$ continuously decreases with redshift (at least in the recent past of the Universe) in a not too unnatural way, then the dimming of distant SNe Ia can be purely due to modified gravity}. Of course, such a scenario may be in conflict with various other local experiments and cosmological observations. However, given the complexity of modified gravity models it is useful to understand in more detail whether those observational constraints can be evaded or whether they need to be reinterpreted in the context of such models if they are to apply to them, as we have shown is the case for for SNe Ia.

Finally, let us mention once more the diversity of modified gravity models. In such scenarios, the internal structure of white dwarfs can be affected and the full modified Einstein equations need to be solved to understand this accurately. This will possibly cause `second-order' effects on the astrophysics of SNe Ia, for example through the different density profile of the white dwarf progenitors. This, coupled to the complicated thermonuclear reactions and radiation transfer inside the ejecta, poses a great challenge for modeling the explosion and aftermath of SNe Ia numerically, let alone analytically. Therefore, much work remains to be done to refine the understanding of the effect of gravity -- both standard and non-standard -- in supernova cosmology.

\acknowledgments
We would like to thank Carlton Baugh, Celine Boehm, Suhail Dhawan, Hans Winther, and John D. Barrow for useful comments. BSW is supported by the U.K. Science and Technology Facilities Council (STFC) research studentship, and thanks the Institute for Computational Cosmology for hosting him when part of the work described in this paper was carried out. BL acknowledges support by the European Research Council (ERC-StG-716532-PUNCA), the ICC's STFC Consolidated Grants (ST/P000541/1, ST/L00075X/1) and Durham University.

\bibliography{main}

\newpage
\begin{appendix}

\section{Full derivation of semi-analytic light curve model}\label{App:LCDerivation}

In this appendix we give a detailed derivation and explanation of the light curve model used in this paper. This appendix is largely a self-contained summary of the works described in Refs. \cite{Arnett1980, Arnett1982, Chatzopoulos2012}.

\subsection{Main Equation}

We will begin with the main equation for the thermodynamics of the supernova, which will allow us to derive an equation for the supernova's luminosity over time -- its light curve. For this the following assumptions and approximations\footnote{We number these as [A1]-[A7], and will mention the numbers where the corresponding assumptions or approximations are actually used in the text below.} are made:
\begin{enumerate}
	\item [{[A1]}] The system is spherically symmetric.
	\item [{[A2]}] The expansion of the supernova ejecta is homologous -- see Eq.~(\ref{eq:homexp}) below -- and the shells of the expanding ejecta do not cross each other.
	\item [{[A3]}] The supernova ejecta gas is dominated by radiation pressure.
	\item [{[A4]}] The supernova ejecta is optically thick with an optical depth $>$ 1. This is known as the diffusion approximation.
	\item [{[A5]}] The effective opacity of the ejecta is constant.
	\item [{[A6]}] Radioactive decay is the only source of energy in the system.
	\item [{[A7]}] The distribution of $^{56}\textup{Ni}$ is concentrated in the centre of the system.
\end{enumerate}
Applying the first law of thermodynamics to the system of expanding supernova ejecta yields:
\begin{equation}
\dot{E} + P \dot{V} = -\frac{\partial L}{\partial m} + \epsilon,
\label{eq:1stlaw}
\end{equation}
where $E=aT^4 V$ is the specific energy density, $P=aT^4/3$ is the pressure [A3], $T$ is the temperature, $V=1/\rho$ the specific volume and $\rho$ the density of the ejecta, $a=4\sigma /c$ is the radiation constant, $\sigma$ is the Stefan-Boltzmann constant, the $\dot{y}$ notation represents the partial derivative with respect to time $\partial y/\partial t$, $L$ is the luminosity output of the system, $m$ the mass and $\epsilon$ is the rate of energy per unit mass added to the system, which is discussed below.

The first term of Eq.~(\ref{eq:1stlaw}), $\dot{E}$, represents the rate of change in specific energy density, and the second term $P\dot{V}$ represents the specific work involved in expanding the ejecta, so that the sum of the rate of change in energy density and the specific work is equal to the sum of the energy per unit mass added to the system (positive) and the luminosity output of the system per unit mass (negative).

The source of energy in this system $\epsilon$ is the radioactive decay of $^{56}\textup{Ni}$ to $^{56}\textup{Co}$ and the subsequent decay of $^{56}\textup{Co}$ to stable $^{56}\textup{\textup{Fe}}$ [A6]:
\begin{equation}
{}^{56}_{28}\textup{Ni} \to {}^{56}_{27}\textup{Co} + {}^{\ 0}_{+1}\mathrm{e}^+ + \gamma \to {}^{56}_{26}\textup{Fe} + 2\ {}^{\ 0}_{+1}\mathrm{e}^+ + \gamma .
\end{equation}

\noindent
The rate of change in the number of $^{56}\textup{Ni}$ nuclei during decay is given by
\begin{equation}
\dot{N}_{\textup{Ni}}(t)=  -\lambda_{\textup{Ni}} N^0_{\textup{Ni}} \mathrm{e}^{-t/\tau_{\textup{Ni}}},
\label{eq:NNidot}
\end{equation}
in which $N^0_{\textup{Ni}}$ is the initial number of $^{56}\textup{Ni}$ nuclei in the system, and $\lambda_{\textup{Ni}}$ and $\tau_{\textup{Ni}}=1/\lambda_{\textup{Ni}}$ are the decay constant and lifetime of $^{56}\textup{Ni}$ respectively. The negative sign shows that the number of $N^0_{\textup{Ni}}$ nuclei is decreasing as the decays occur. Thus the actual rate of $^{56}\textup{Ni}$ to $^{56}\textup{Co}$ decays is
\begin{equation}
\dot{N}_{\textup{Ni}, decay}(t) = \lambda_{\textup{Ni}} N^0_{\textup{Ni}} \mathrm{e}^{-t/\tau_{\textup{Ni}}} .
\label{eq:Nidec}
\end{equation}
The rate of change in the number of $^{56}\textup{Co}$ nuclei is given by
\begin{equation}
\dot{N}_{\textup{Co}}(t)= \lambda_{\textup{Ni}} N^0_{\textup{Ni}} \mathrm{e}^{-t/\tau_{\textup{Ni}}} - \lambda_{\textup{Co}} N_{\textup{Co}},
\label{eq:NCodot}
\end{equation}

\noindent
where the first, positive term on the RHS represents the production of $^{56}\textup{Co}$ nuclei by the decay of $^{56}\textup{Ni}$, and the second, negative term represents the decay of $^{56}\textup{Co}$ nuclei. Solving this equation for $N_{\textup{Co}}$ yields
\begin{equation}
N_{\textup{Co}} = \frac{\lambda_{\textup{Ni}}}{\lambda_{\textup{Ni}}-\lambda_{\textup{Co}}} N^0_{\textup{Ni}} \left(\mathrm{e}^{-t/\tau_{\textup{Co}}} - \mathrm{e}^{-t/\tau_{\textup{Ni}}}\right).
\label{eq:NCo}
\end{equation}
The rate of $^{56}\textup{Co}$ to $^{56}\textup{\textup{Fe}}$ decays is given by
\begin{equation}
\dot{N}_{\textup{Co}, decay}(t) =  \lambda_{\textup{Co}} \frac{\lambda_{\textup{Ni}}}{\lambda_{\textup{Ni}}-\lambda_{\textup{Co}}} N^0_{\textup{Ni}} \left(\mathrm{e}^{-t/\tau_{\textup{Co}}} - \mathrm{e}^{-t/\tau_{\textup{Ni}}}\right),
\label{eq:Codec}
\end{equation}
\noindent
where $\dot{N}_{\textup{Co}, decay}(t)$ is the number of $^{56}\textup{Co}$ decays per second at time $t$. The total rate of energy produced, $\dot{W}$, is then
\begin{equation}
\dot{W}(t) = W_{\textup{Ni}} \dot{N}_{\textup{Ni}, decay}(t) + W_{\textup{Co}} \dot{N}_{\textup{Co}, decay}(t),
\label{eq:Wrate}
\end{equation}
\noindent
where $W_{\textup{Ni}}$ and $W_{\textup{Co}}$ are the energies released in a single $^{56}\textup{Ni}$ decay  and $^{56}\textup{Co}$ decay respectively. The total rate of energy produced per unit mass of $^{56}\textup{Ni}$, $\epsilon$, is given by
\begin{equation}
\epsilon = \frac{\dot{W}(t)}{M_{\textup{Ni}}} = \frac{\dot{W}(t)}{m_{\textup{Ni}}N^0_{\textup{Ni}}},
\label{eq:eps}
\end{equation}
\noindent
where $M_{\textup{Ni}}$ is the initial mass of $^{56}\textup{Ni}$ and $m_{\textup{Ni}}$ the mass per $^{56}\textup{Ni}$ nuclei. Substituting Eqs.~(\ref{eq:Nidec}) and (\ref{eq:Codec}) into Eq.~(\ref{eq:Wrate}), Eq.~(\ref{eq:eps}) becomes
\begin{equation}
\epsilon = \frac{W_{\textup{Ni}}\lambda_{\textup{Ni}}}{m_{\textup{Ni}}} \mathrm{e}^{-t/\tau_{\textup{Ni}}} + \frac{W_{\textup{Co}}\lambda_{\textup{Co}}\lambda_{\textup{Ni}}}{m_{\textup{Ni}}(\lambda_{\textup{Ni}}-\lambda_{\textup{Co}})} \left(\mathrm{e}^{-t/\tau_{\textup{Co}}} - \mathrm{e}^{-t/\tau_{\textup{Ni}}}\right).
\label{eq:eps2}
\end{equation}
Defining
\begin{equation*}
\epsilon_{\textup{Ni}} \equiv \frac{W_{\textup{Ni}}\lambda_{\textup{Ni}}}{m_{\textup{Ni}}},
\label{eq:epsNi}
\end{equation*}
and
\begin{equation*}
\epsilon_{\textup{Co}} \equiv \frac{W_{\textup{Co}}\lambda_{\textup{Co}}\lambda_{\textup{Ni}}}{m_{\textup{Ni}}(\lambda_{\textup{Ni}}-\lambda_{\textup{Co}})},
\label{eq:epsCi}
\end{equation*}
Eq.~(\ref{eq:eps2}) becomes
\begin{equation}
\epsilon = (\epsilon_{\textup{Ni}}-\epsilon_{\textup{Co}}) \mathrm{e}^{-t/\tau_{\textup{Ni}}} + \epsilon_{\textup{Co}} \mathrm{e}^{-t/\tau_{\textup{Co}}}.
\label{eq:eps3}
\end{equation}

\subsection{Terms in main equation}

Because we assume spherical symmetry [A1] the quantities in the main equation depend only on radial coordinate $r$. In the diffusion approximation [A4], the luminosity of a shell of the ejecta at radius $r$ is related to the temperature of that shell by:
\begin{equation}
L = -4\pi r^2 \frac{\Gamma c}{3} \frac{\partial aT^4}{\partial r},
\label{eq:luminosity}
\end{equation}
where $r$ is the radial coordinate, $\Gamma=1/\rho \kappa$ is the mean free path in the ejecta, and $\kappa\equiv\kappa(r)$ is the opacity of the ejecta.

Shortly after the initial supernova explosion, the expansion of the ejecta should become homologous [A2] such that the radius depends only on time $t$:
\begin{equation}
R(t) = R_0 + v_{\mathrm{sc}}t,
\label{eq:homexp}
\end{equation}
where the radial extent of the surface of the ejecta at time $t$, $R(t)$, has advanced constantly at a scale velocity $v_{\mathrm{sc}}$ from its initial position $R_0$. We assume that the shells of expanding ejecta do not cross each other [A2], which we model by writing the velocity of sub-surface layers of the ejecta as
\begin{equation}
v(x) = xv_{\mathrm{sc}},
\label{eq:velo}
\end{equation}
where a change of coordinate to dimensionless radius $x=r(t)/R(t)$ has been carried out such that $x=[0, 1]$.

We separate the time and spatial dependence of the ejecta density profile:
\begin{equation}
\rho(x,t)=\rho_{00} \eta(x) {\left[\frac{R(t)}{R_0}\right]}^{-3},
\label{eq:dens}
\end{equation}
where $\eta(x)$ is the dimensionless, time independent run of density, and $\rho_{00}$ is the initial central density. Using the expression $V=1/\rho$ we can write
\begin{equation}
V(x,t)=\frac{V_{00}}{\eta(x)} {\left[\frac{R(t)}{R_0}\right]}^3,
\label{eq:vol}
\end{equation}
where $V_{00}=1/\rho_{00}$ is the initial specific volume at $r=0$. Given that $V \propto R^3$ we can then write
\begin{equation}
\frac{\dot{V}}{V}=\frac{3\dot{R}}{R}=\frac{3v_{\mathrm{sc}}}{R},
\label{eq:VRdiff}
\end{equation}
where the final equality comes from differentiating Eq.~(\ref{eq:homexp}). Applying a similar separation of variables into space and time to the ejecta temperature yields
\begin{equation}
T(r,t)^4=\psi(x) \phi(t) T^4_{00} {\left[\frac{R_0}{R(t)}\right]}^4.
\label{eq:temp}
\end{equation}
where the spatial dependence is given by $\psi(x)$, and the time dependence is not purely due to expansion effects through $R(t)$ but also an additional term $\phi(t)$ which encapsulates the temperature changing over time due to energy gain/loss. $T_{00}$ again gives the initial central temperature.

Substituting Eq.~(\ref{eq:luminosity}) into Eq.~(\ref{eq:1stlaw}) yields
\begin{equation}
4T^4 \left(\frac{\dot{T}}{T}+\frac{\dot{V}}{3V}\right)=\frac{1}{r^2} \frac{\partial}{\partial r} \left[\frac{c}{3\kappa\rho}r^2\frac{\partial T^4}{\partial r}\right] + \frac{\epsilon}{aV}
\label{eq:4t4}
\end{equation}
We get
\begin{widetext}
\begin{equation}
\frac{R_0}{R(t)} \frac{\dot{\phi}(t)}{\phi(t)} - \frac{1}{aT_{00}^4V_{00}}\frac{b(x)}{\phi(t)} \left[ (\epsilon_{\rm Ni} - \epsilon_{\rm Co} ) \exp^{-\frac{t}{t_{\rm Ni}}} + \epsilon_{\rm Co}\exp^{-\frac{t}{t_{\rm Co}}} \right] = -\frac{\alpha(x) c}{3R_0^2 \kappa(0)\rho_{00}},
\label{eq:longtrans}
\end{equation}
\end{widetext}
where we have assumed $\kappa(x)=\kappa(0)$ [A5] and defined
\begin{equation}
\alpha(x) \equiv -\frac{1}{x^2\psi} \frac{{\rm d}}{{\rm d}x} \left(\frac{x^2}{\eta(x)}\frac{{\rm d}\psi}{{\rm d}x}\right),
\label{eq:alpha}
\end{equation}
\begin{equation}
b(x) \equiv \frac{\xi(x)\eta(x)}{\psi(x)}.
\label{eq:b}
\end{equation}
Note that to identify which radial layer different amounts of energy are produced in a radial dependence is added to $\epsilon$ based on the distribution of $^{56}\textup{Ni}$ given by $\xi(x)$:
\begin{equation}
\epsilon(x, t) = \xi(x) \left[ (\epsilon_{\textup{Ni}}-\epsilon_{\textup{Co}}) \mathrm{e}^{-t/\tau_{\textup{Ni}}} + \epsilon_{\textup{Co}} \mathrm{e}^{-t/\tau_{\textup{Co}}} \right].
\label{eq:eps4}
\end{equation}
Here, $b$ is approximately constant for any $x$ under the assumption that the $^{56}\textup{Ni}$ is concentrated in the centre of the ejecta [A7]. Through the definition of $\alpha$ in Eq.~(\ref{eq:alpha}), $\alpha$ depends only on $x$. Meanwhile, with $b$ as a constant the LHS of Eq.~(\ref{eq:longtrans}) implies $\alpha$ is a function of $t$ only, therefore $\alpha$ must be a constant. Because $\alpha$ is a constant, we can define a constant $\tau_0$:
\begin{equation}
\tau_0 \equiv \frac{3R^2_0 \rho_{00} \kappa(0)}{\alpha c},
\label{eq:tau0}
\end{equation}
and then rewrite Eq.~(\ref{eq:longtrans}) as
\begin{equation}
-\frac{\phi}{\tau_0} = \frac{R_0}{R(t)} \frac{{\rm d}\phi}{{\rm d}t} - \left[\frac{b}{aT^4_{00}V_{00}}\right] \left[ (\epsilon_{\textup{Ni}}-\epsilon_{\textup{Co}}) \mathrm{e}^{-t/\tau_{\textup{Ni}}} + \epsilon_{\textup{Co}} \mathrm{e}^{-t/\tau_{\textup{Co}}} \right],\nonumber
\label{eq:alphatau0}
\end{equation}
which more neatly can be written as
\begin{equation}
\dot{\phi} + \frac{\phi R(t)}{R_0 \tau_0} = \frac{\tilde{\epsilon}R(t)}{R_0},
\label{eq:phidot2}
\end{equation}
with
\begin{equation}
\tilde{\epsilon} \equiv \frac{b}{aT^4_{00}V_{00}} \left[ (\epsilon_{\textup{Ni}}-\epsilon_{\textup{Co}}) \mathrm{e}^{-t/\tau_{\textup{Ni}}} + \epsilon_{\textup{Co}} \mathrm{e}^{-t/\tau_{\textup{Co}}} \right].
\label{eq:epssquig}
\end{equation}
Eq.~(\ref{eq:phidot2}) is the key equation we need to solve to compute the light curve of the SN Ia. In the next subsection we will do this explicitly.

\subsection{Light curve solutions}

Let $\dot{u}=R(t)/R_0\tau_0$, and substitute in Eq.~(\ref{eq:homexp}) to get
\begin{equation}
\dot{u}=\frac{1}{\tau_0} + \frac{v_{\mathrm{sc}}t}{R_0\tau_0},
\label{eq:udot}
\end{equation}
and therefore $u$ is given by
\begin{equation}
u = \frac{t}{\tau_0} + \frac{v_{\mathrm{sc}}t^2}{2R_0\tau_0} = \frac{t}{\tau_0} + \frac{t^2}{2\tau_h\tau_0} = \frac{t}{\tau_0} + \frac{t^2}{\tau^2_m},
\label{eq:u}
\end{equation}
where $\tau_h\equiv R_0/v_{\mathrm{sc}}$ is the expansion timescale and $\tau^2_m\equiv 2\tau_0\tau_h$ is the light curve timescale. Substituting Eq.~(\ref{eq:udot}) into Eq.~(\ref{eq:phidot2}) yields
\begin{equation}
\frac{\tilde{\epsilon}R(t)}{R_0} = \dot{\phi} +\phi\dot{u} = \mathrm{e}^{-u} \frac{{\rm d}}{{\rm d}t}\left(\phi \mathrm{e}^{u}\right),
\label{eq:phidot3}
\end{equation}
the solution to which, $\phi(t)$, is directly related to the light curve of the SN Ia. Eq.~(\ref{eq:phidot3}) depends on $\tilde{\epsilon}$, which itself depends on constants such as $b$, $V_{00}$ and $T_{00}$. It is useful to express these constants in terms of more physical quantities. For this, we first define
\begin{equation}
I_M \equiv \int^{1}_{0} \eta(x) x^2 {\rm d}x = \frac{V_{00}M}{4\pi R^3_0},\nonumber
\label{eq:IM}
\end{equation}
where $M$ is the total ejecta mass, and substitute into Eq.~(\ref{eq:tau0}) to get
\begin{equation}
\tau_0 = \frac{3\kappa M}{4\pi\alpha c I_M R_0} = \frac{\kappa M}{\beta c R_0},
\label{eq:tau02}
\end{equation}
in which $\beta$ is defined as $\beta \equiv 4\pi\alpha I_M/3$. Reference~\cite{Arnett1980} discuses solutions to Eq.~(\ref{eq:alpha}) and the corresponding boundary conditions at length, and finds $\beta$ to be approximately a constant $\beta \approx 13.7$ for a variety of different density distributions. This value shall be used hereafter.
On the other hand, the mass of $^{56}\textup{Ni}$ initially present in the ejecta is given by
\begin{equation}
M_{\textup{Ni}} = \frac{4\pi R^3_0}{V_{00}} \int^{1}_{0} \xi(x) \eta(x) x^2 {\rm d}x,
\label{eq:MNi0}
\end{equation}
and the total thermal energy content is given by
\begin{equation}
E_{Th}(t) = \int aT^4 {\rm d}V = 4\pi R^3_0 a T^4_{00} \phi(t) \frac{R_0}{R(t)}I_{Th},
\label{eq:ETh2}
\end{equation}
where we have used Eq.~(\ref{eq:temp}) and defined
\begin{equation}
I_{Th} \equiv \int^1_0 \psi(x) x^2 {\rm d}x.
\label{eq:ITh}
\end{equation}
Inserting Eqs.~(\ref{eq:IM}), (\ref{eq:MNi0}) and (\ref{eq:ITh}) into Eq.~(\ref{eq:b}) yields a new expression for $b$:
\begin{equation}
b = \frac{M_{\textup{Ni}} I_M}{M I_{Th}}.
\label{eq:b2}
\end{equation}
Similarly, we can use Eq.~(\ref{eq:IM}) to rewrite $V_{00}$ as
\begin{equation}
V_{00} = \frac{4\pi R^3_0 I_M}{M},\nonumber
\label{eq:V00}
\end{equation}
and use Eq.~(\ref{eq:ETh2}) to rewrite $aT^4_{00}$:
\begin{equation}
aT^4_{00} = \frac{R(t) E_{Th}(t)}{4\pi R^4_0 I_{Th} \phi(t)}.\nonumber
\label{eq:aT400}
\end{equation}
Substituting these, and Eq.~(\ref{eq:b2}), into Eq.~(\ref{eq:epssquig}), we get
	\begin{equation}
\tilde{\epsilon}(t) = \frac{M_{\textup{Ni}} R_0 \phi(t)}{E_{Th}(t)R(t)} \left[ (\epsilon_{\textup{Ni}}-\epsilon_{\textup{Co}}) \mathrm{e}^{-t/\tau_{\textup{Ni}}} + \epsilon_{\textup{Co}} \mathrm{e}^{-t/\tau_{\textup{Co}}} \right],
\label{eq:epssquig2}
\end{equation}
which at $t=0$ becomes $\tilde{\epsilon}(0)=\frac{M_{\textup{Ni}}}{E_{Th}(0)}\epsilon_{\rm Ni}$, in which we have used $R(0)=R_0$ and $\phi(0)=1$. Therefore, Eq.~(\ref{eq:phidot3}) can be rewritten in terms of more physically meaningful quantities as
\begin{widetext}
\begin{equation}
\frac{{\rm d}}{{\rm d}t}\left(\mathrm{e}^{\frac{t}{\tau_0} + \frac{t^2}{\tau^2_m}} \phi\right) = \frac{M_{\textup{Ni}}\tau_0}{E_{Th}(0)} \left( \frac{1}{\tau_0} + \frac{v_{\mathrm{sc}}t}{R_0\tau_0} \right) \mathrm{e}^{\frac{t}{\tau_0} + \frac{t^2}{\tau^2_m}} \left[ (\epsilon_{\textup{Ni}}-\epsilon_{\textup{Co}}) \mathrm{e}^{-t/\tau_{\textup{Ni}}} + \epsilon_{\textup{Co}} \mathrm{e}^{-t/\tau_{\textup{Co}}}\right],
\label{eq:phi1}
\end{equation}
\end{widetext}
which can be solved as
\begin{widetext}
\begin{multline}
\phi(t) = \frac{M_{\textup{Ni}}\tau_0}{E_{Th}(0)} \mathrm{e}^{ -\left(\frac{2 R_0 t}{v_{\mathrm{sc}} \tau^2_m} + \frac{t^2}{\tau^2_m}\right) } \bigg[ (\epsilon_{\textup{Ni}}-\epsilon_{\textup{Co}}) \int^{t}_{0} \left(\frac{R_0}{v_{\mathrm{sc}}\tau_m} + \frac{t^{\prime}}{\tau_m}\right) \mathrm{e}^{\left( \frac{t^{\prime 2}}{\tau^2_m} + \frac{2R_0 t^{\prime}}{v_{\mathrm{sc}}\tau^2_m} \right)} \mathrm{e}^{-t^{\prime}/\tau_{\textup{Ni}}} {\rm d}t^{\prime} \\
 + \epsilon_{\textup{Co}} \int^{t}_{0} \left(\frac{R_0}{v_{\mathrm{sc}}\tau_m} + \frac{t^{\prime}}{\tau_m}\right) \mathrm{e}^{\left( \frac{t^{\prime 2}}{\tau^2_m} + \frac{2R_0 t^{\prime}}{v_{\mathrm{sc}}\tau^2_m} \right)} \mathrm{e}^{-t^{\prime}/\tau_{\textup{Co}}} {\rm d}t^{\prime} \bigg].
\label{eq:phi2}
\end{multline}
\end{widetext}
Finally, $\phi(t)$ can be related to the luminosity output of the SN Ia by inserting Eq.~(\ref{eq:temp}) into Eq.~(\ref{eq:luminosity}), which gives
\begin{equation}
L(x,t) = -\frac{16\pi^2 a c T^4_{00} R^4_0 I_M}{3\kappa M} \phi(t) \left( -\frac{x^2}{\eta(x)}\frac{{\rm d}\psi}{{\rm d}x} \right),
\label{eq:luminosity2}
\end{equation}
in which we have used $\Gamma=1/\kappa\rho$, $\rho=1/V$, and Eq.~(\ref{eq:vol}) to write
\begin{equation}
\Gamma = \frac{V_{00}R^3(t)}{\kappa \eta(x) R^3_0} = \frac{4\pi R^3(t) I_M}{\kappa M \eta(x)}
\label{eq:Gamma}
\end{equation}
where in the last step we have used Eq.~(\ref{eq:V00}).

The surface luminosity is given by $x=1$ (or $r=R$), as
\begin{equation}
L(1,t) = -\frac{16\pi^2 a c T^4_{00} R^4_0 I_M}{3\kappa M} \phi(t) \left[-\frac{x^2}{\eta(x)}\frac{{\rm d}\psi}{{\rm d}x} \right]_{x=1}.
\label{eq:luminosity3}
\end{equation}
To remove the spatial derivative in the brackets, we rearrange Eq.~(\ref{eq:alpha}), which leads to
\begin{equation}
x^2 \psi(x) \alpha = \frac{{\rm d}}{{\rm d}x} \left(-\frac{x^2}{\eta(x)}\frac{{\rm d}\psi}{{\rm d}x}\right).\nonumber
\end{equation}
Integrating both sides of this equation between $x=0$ and $x=1$, and comparing to Eq.~(\ref{eq:ITh}), we find
\begin{equation}
\left[-\frac{x^2}{\eta(x)}\frac{{\rm d}\psi}{{\rm d}x} \right]_{x=1} = \alpha \int^1_0 x^2 \psi(x){\rm d}x = \alpha I_{Th}.\nonumber
\end{equation}
Substituting this into Eq.~(\ref{eq:luminosity3}) yields
\begin{equation}
L(1,t) = -\frac{16\pi^2 a c T^4_{00} R^4_0 I_M \alpha I_{Th}}{3\kappa M} \phi(t).
\label{eq:luminosity4}
\end{equation}
Finally, using Eqs.~(\ref{eq:tau02}), (\ref{eq:ETh2}), (\ref{eq:phi2}) and the definition of $\beta$, Eq.~(\ref{eq:luminosity4}) becomes
\begin{widetext}
\begin{multline}
L(1,t)=\frac{2M_{\textup{Ni}}}{\tau_m} \mathrm{e}^{ -\left(\frac{2 R_0 t}{v_{\mathrm{sc}} \tau^2_m} + \frac{t^2}{\tau^2_m}\right) } \bigg[ (\epsilon_{\textup{Ni}}-\epsilon_{\textup{Co}}) \int^{t}_{0} \left(\frac{R_0}{v_{\mathrm{sc}}\tau_m} + \frac{t^{\prime}}{\tau_m}\right) \mathrm{e}^{\left( \frac{t^{\prime 2}}{\tau^2_m} + \frac{2R_0 t^{\prime}}{v_{\mathrm{sc}}\tau^2_m} \right)} \mathrm{e}^{-t^{\prime}/\tau_{\textup{Ni}}} {\rm d}t^{\prime}\\
 + \epsilon_{\textup{Co}} \int^{t}_{0} \left(\frac{R_0}{v_{\mathrm{sc}}\tau_m} + \frac{t^{\prime}}{\tau_m}\right) \mathrm{e}^{\left( \frac{t^{\prime 2}}{\tau^2_m} + \frac{2R_0 t^{\prime}}{v_{\mathrm{sc}}\tau^2_m} \right)} \mathrm{e}^{-t^{\prime}/\tau_{\textup{Co}}} {\rm d}t^{\prime}\bigg].
\label{eq:luminosity5}
\end{multline}
\end{widetext}
As we want to produce ultraviolet+optical+infrared (UVOIR) light curves, a factor accounting for the possibility of gamma ray leakage, where gamma ray photons escape directly through the ejecta without interaction, should be included such that
\begin{widetext}
\begin{multline}
L(1,t)=\frac{2M_{\textup{Ni}}}{\tau_m} \mathrm{e}^{ -\left(\frac{2 R_0 t}{v_{\mathrm{sc}} \tau^2_m} + \frac{t^2}{\tau^2_m}\right) } \bigg[ (\epsilon_{\textup{Ni}}-\epsilon_{\textup{Co}}) \int^{t}_{0} \left(\frac{R_0}{v_{\mathrm{sc}}\tau_m} + \frac{t^{\prime}}{\tau_m}\right) \mathrm{e}^{\left( \frac{t^{\prime 2}}{\tau^2_m} + \frac{2R_0 t^{\prime}}{v_{\mathrm{sc}}\tau^2_m} \right)} \mathrm{e}^{-t^{\prime}/\tau_{\textup{Ni}}} {\rm d}t^{\prime} \\
 + \epsilon_{\textup{Co}} \int^{t}_{0} \left(\frac{R_0}{v_{\mathrm{sc}}\tau_m} + \frac{t^{\prime}}{\tau_m}\right) \mathrm{e}^{\left( \frac{t^{\prime 2}}{\tau^2_m} + \frac{2R_0 t^{\prime}}{v_{\mathrm{sc}}\tau^2_m} \right)} \mathrm{e}^{-t^{\prime}/\tau_{\textup{Co}}} {\rm d}t^{\prime} \bigg] \left(1-\mathrm{e}^{(t_0/t)^2}\right),
\label{eq:luminosity6}
\end{multline}
\end{widetext}

\noindent
where $t_0 = (9\kappa_{\gamma}/2\pi E_{\mathrm{K}})^{1/2}$ is the timescale for gamma ray leakage, $\kappa_{\gamma}$ is the ejecta's gamma ray opacity, and $E_{\mathrm{K}}$ is the kinetic energy in the supernova explosion. Eq.~(\ref{eq:luminosity6}) is the equation to compute UVOIR light curves used throughout the paper.

\section{Derivation of the Chandrasekhar mass} \label{App:MChDerivation}

The calculation of the Chandrasekhar mass $M_{\rm Ch}$ is now standard textbook material (see for example \cite{MChDerivation}). Here we include a derivation to make the paper self-contained, and to highlight some physics that is relevant for the discussion of this paper.

We start from the equation of hydrostatic equilibrium for a spherically symmetrical stellar fluid in Newtonian gravity:
\begin{equation}
\frac{{\rm d}P}{{\rm d}r} = - \rho(r)\frac{GM(r)}{r^2},
\label{eq:hydrostat}
\end{equation}
where $r$ is the radial coordinate, $P$ is the pressure of the fluid, $\rho$ is the density, and $M(r)$ is the mass within a sphere of radius $r$, given by ${\rm d}M=4\pi r^2\rho(r){\rm d}r$ or
\begin{equation}
M(r) = \int^{r}_{0} 4\pi r^2\rho(r){\rm d}r.
\label{eq:M(r)}
\end{equation}
Rearranging Eq.~(\ref{eq:hydrostat}) for $M(r)$ and differentiating with respect to radius $r$ yields
\begin{equation}
\frac{{\rm d}M}{{\rm d}r} = - \frac{1}{G} \frac{{\rm d}}{{\rm d}r} \left( \frac{r^2}{\rho} \frac{{\rm d}P}{{\rm d}r}\right).\nonumber
\label{eq:dMdr}
\end{equation}
Comparing this with Eq.~(\ref{eq:M(r)}), we get the familiar equation
\begin{equation}
\frac{1}{r^{2}} \frac{{\rm d}}{{\rm d}r} \left( \frac{r^{2}}{\rho} \frac{{\rm d}P}{{\rm d}r} \right) = -4 \pi G \rho.
\label{eq:stellarstate}
\end{equation}
To solve this equation, an equation of state relating $P$ and $\rho$ is required. For the degenerate material under high amounts of compression inside a white dwarf, the electrons have a large energy due to the electron degeneracy pressure, and so have a velocity approaching that of the speed of light. Therefore, the white dwarf material is best described as a relativistic Fermi gas with an equation of state given by
\begin{equation}
P = \frac{\hbar c}{12 \pi^{2}} \left( \frac{3 \pi^{2} \rho}{m_{\mathrm{N}} \mu} \right)^{4/3},
\label{eq:EoS}
\end{equation}
in which $\hbar$ is the reduced Planck constant, $m_{\mathrm{N}}$ is the nucleon mass, and $\mu=\left\langle A/Z \right\rangle$ is the average mass number per nuclear charge with $\mu \approx2$ for the $^{12}\mathrm{C}$ and $^{16}\mathrm{O}$ that make up the majority of the white dwarf. This equation of state is of the form of a polytrope
\begin{equation}
P = K \rho^{\gamma},\nonumber
\label{eq:polytrope}
\end{equation}
with $K\equiv\hbar c / 12 {\pi}^2 \times {\left( 3 {\pi}^2/m_{\mathrm{N}} \mu \right)}^{4/3}$ and $\gamma=4/3$. Using this equation of state, Eq.~(\ref{eq:stellarstate}) can be rewritten as
\begin{equation}
\left[ \frac{n+1}{4 \pi G} K \lambda^{(1-n)/n} \right] \frac{1}{r^2} \frac{{\rm d}}{{\rm d}r}\left( r^2 \frac{{\rm d}\Theta}{{\rm d}r} \right) = -\Theta^n
\label{eq:polytropestate3}
\end{equation}
where we have defined $\rho \equiv \lambda \Theta^n$ and $\gamma \equiv \frac{n+1}{n}$. This equation can be made dimensionless by introducing a  radial variable $y \equiv r/\alpha$ where $\alpha \equiv \sqrt{(n+1)K \lambda^{(1-n)/n}/4 \pi G}$, yielding
\begin{equation}
\frac{1}{y^2} \frac{{\rm d}}{{\rm d}y} \left(y^2 \frac{{\rm d}\Theta}{{\rm d}y} \right) = - \Theta^n.
\label{eq:LaneEm}
\end{equation}
This is the Lane-Emden equation for polytropes in hydrostatic equilibrium. As a second-order ordinary differential equation it requires two boundary conditions to complete it. First, we can define the central density as $\rho(r=0)=\rho_{\mathrm{c}} \equiv \lambda$, which gives
\begin{equation}
\rho_{\mathrm{c}} = \lambda \Theta^n(y=0) = \lambda \Rightarrow \Theta (y=0) = 1.
\label{eq:bound1}
\end{equation}

\noindent
Then, since $M(r=0)=0$ physically, we have
\begin{equation}
{\frac{{\rm d}P}{{\rm d}r}\Big|}_{r=0} = - \rho_{\mathrm{c}} \frac{GM(r=0)}{r^2} = 0.\nonumber
\end{equation}
For the polytropic equation of state introduced above, this is equivalent to
\begin{equation}
{\frac{{\rm d}P}{{\rm d}r}\Big|}_{r=0} = \gamma K \rho_{\mathrm{c}}^{\gamma - 1} {\frac{{\rm d}\rho}{{\rm d}r}\Big|}_{r=0} = 0,\nonumber
\end{equation}
and can be re-expressed in terms of the dimensionless quantities $\Theta$ and $y$ as
\begin{equation}
\left[\frac{{\rm d}\Theta}{{\rm d}y}\right]_{y=0} = 0.
\label{eq:bound2}
\end{equation}
The outer radius of the star, $R$, corresponds to the point $y=y_1$ at which the density drops to zero, i.e., $\rho(R) = \Theta(y_1) = 0$. Using $\rho \equiv \lambda \Theta^n$ and $y \equiv r/\alpha$, the equation for the total mass of the star, Eq.~(\ref{eq:M(r)}), becomes
\begin{equation}
M = 4 \pi \lambda \alpha^3 \int^{y_1}_{0} y^2 \Theta^n {\rm d}y,
\label{eq:Mint1}
\end{equation}
and then using Eq.~(\ref{eq:LaneEm}) this becomes
\begin{equation}
M = 4\pi\lambda\alpha^3 \int^{y_1}_{0} -\frac{{\rm d}}{{\rm d}y} \left( y^2 \frac{{\rm d}\Theta}{{\rm d}y} \right) {\rm d}y = 4 \pi \lambda \alpha^3 \left[-y^2 \frac{{\rm d}\Theta}{{\rm d}y}\right]_{y_1},
\label{eq:Mint3}
\end{equation}

Therefore, calculating the total mass requires the value of $y_1$ to be computed by solving the Lane-Emden equation. Recall now that the specific case of a white dwarf yields a polytrope with $\gamma=4/3$, which corresponds to a polytropic index $n=3$. The corresponding Lane-Emden equation is
\begin{equation}
\frac{1}{y^2} \frac{{\rm d}}{{\rm d}y} \left(y^2 \frac{{\rm d}\Theta}{{\rm d}y} \right) + \Theta^3 = 0,
\label{eq:LaneEm3}
\end{equation}
which can be solved numerically to find $y_1$ at which $\Theta(y_1) = 0$. Doing so yields a value $y_1=6.89685$ and $-y^2 \Theta^{\prime}(y_1)=2.01824$. Using the definition of $\alpha$:
\begin{equation}
4 \pi \lambda \alpha^3 = 4 \pi \lambda {\left( \frac{(n+1)K \lambda^{(1-n)/n}}{4 \pi G} \right)}^{3/2},
\end{equation}
and plugging in $n=3$ gives
\begin{equation}
4 \pi \lambda \alpha^3 = 4 \pi \lambda \left( \frac{4K \lambda^{-2/3} }{4 \pi G } \right) ^{3/2}  = 4 \pi { \left( \frac{K}{\pi G} \right) }^{3/2}.
\label{eq:lamalph}
\end{equation}
In the specific case of a white dwarf, $K=\hbar c / 12{\pi}^2 \times {\left( 3 {\pi}^2/ \mu \/ m_{\mathrm{N}}  \right)}^{4/3}$, so Eq.~(\ref{eq:lamalph}) becomes
\begin{equation}
4 \pi \lambda \alpha^3 = 4 \pi {\left( \frac{K}{\pi G} \right)}^{3/2} = \frac{\sqrt{3 \pi}}{2} {\left( \frac{\hbar c}{G} \right)}^{3/2} \frac{1}{( \mu m_{\mathrm{N}})^2}
\label{eq:lamalph3}.
\end{equation}
Inserting Eq.~(\ref{eq:lamalph3}) into Eq.~(\ref{eq:Mint3}) yields
\begin{equation}
M_{\rm Ch} = \frac{\sqrt{3 \pi}}{2} {\left( \frac{\hbar c}{G} \right)}^{3/2} \frac{1}{( \mu m_{\mathrm{N}} )^2} \left[- y^2 \frac{{\rm d}\Theta}{{\rm }dy} \right]_{y_1}
\label{eq:Mint4}
\end{equation}
Plugging $y_1(n=3)=6.89685$ and $-y^2 \Theta^{\prime}(y_1)=2.01824$ into Eq.~(\ref{eq:Mint4}) results in $M_{\mathrm{Ch}}=1.44M_{\odot}$. In our case, note that, in addition to the numerical value, $M_{\rm Ch}$ has a specific dependence on $G$: $M_{\rm Ch}\propto G^{-3/2}$. Physically, a smaller value of $G$ means that gravity is weaker, so that the electron degeneracy pressure can support a larger mass.

\end{appendix}
\end{document}